# Internal gravity waves in a stratified medium of non-uniform depth

**Vitaly V. Bulatov, Yury V. Vladimirov**
**Institute for Problems in Mechanics**
**Russian Academy of Sciences**
**Pr. Vernadskogo 101 - 1, Moscow, 119526, Russia**
**bulatov@index-xx.ru**

**Abstract**
   The problem of reconstructing non-harmonic internal gravity wave packets generated by a source moving in a stratified ocean is considered. The uniform asymptotic form of the internal gravity waves field generated by a source moving above the smoothly varying floor is constructed. The solution is proposed in terms of wave modes, propagating independently at the adiabatic approximation, and described as a non-integral degree series of a small parameter characterizing the stratified medium. A specific form of the wave packets, which can be parameterized in terms of model functions (Airy functions), depends on a local behavior of the dispersion curves of individual wave mode. A modified space-time ray method was proposed, which belongs to the class of geometrical optics methods. The key point of the proposed technique is the possibility to derive the asymptotic representation of the solution in terms of a non-integral degree series of the some small parameter.

## 1. Introduction

   As is well known, an essential influence of the propaganda of internal gravity waves in stratified natural media (ocean, atmosphere) is  caused by the horizontal inhomogeneity and nonstationarity of these media. To the most typical horizontal inhomogeneitities of a real ocean one can refer the modification of the relief of the bottom, and inhomogeneity of the density field, and the variability of the mean flows. One can obtain an exact analytic solution of this problem (for instance, by using the method of separation of variables) only id the distribution of density and the shape of the bottom are described by rather simple model functions. If the shape of the bottom and the stratification are arbitrary, then one can construct only asymptotic representation of the solution in the near and far zones; however, to describe the field of internal waves between these zones, one needs an accurate numerical solution of the problem .
   Using asymptotic methods, one can consider a wide class of interesting physical problems, including problems concerning the propagation of nonharmonic wave packets of internal gravity waves in diverse nonhomogeneous stratified media under the assumption that the modification of the parameters of a vertically stratified medium are slow in the horizontal direction. From the general  point of view, problems of this kind can be studied in the framework of a combination of the adiabatic and semiclassical approximations or by using close approach, for axample, ray expansions. In particular, the asymptotic solutions of diverse dynamical problems can be described by using the Maslov canonical operator,  which determines the asymptotic behavior of the solution, including the case of neighborhoods of singular sets composed of  focal points, caustics, etc. The specific form  of the wave packet can be finally expressed by using some special functions, slay, in terms of oscillating  exponentials, Airy function, Fresnel integral, Pearcey-type integral, etc.
   The above approaches are quite general and, in principle, enable one to solve a broad spectrum of  problems from the mathematical point of view; however, the jproblem of their practical applications and, in particular, of the visualization of the corresponding asymptotic formulas based on the Maslov canonical operator is still far from completion, and in some specific problems to find the asymptotic behavior whose computer realization using software of *Mathematica* type is rather simple. In this paper, using the approaches developed in [1], we construct and numerically realize asymptotic solutions of the problem which is formulated as follows.

The internal gravity waves are the oscillations of a stratified medium in the gravity force field. The stratified medium is such a medium where the density increases with the depth. Suppose that a volume element of the medium is not at the equilibrium, for example it could be displaced upward, then it will be heavier than the surrounding medium and therefore Archimedean forces will make it move back to the equilibrium. The essential parameter of any oscillating system is the frequency. It is determined by the correlation of two factors: returning forces which return the perturbed system towards its equilibrium and the inertial forces. For the internal gravity waves the returning forces are proportional to the vertical gradient of the fluid's density and the inertial ones are proportional to the density itself. For the characteristic frequency of the gravity waves oscillations we have the following expression: $N^2(z) = -\frac{g}{\rho(z)}\frac{d\rho(z)}{dz}$ [1, Chap.1]. This frequency is usually called by the Brunt-Vaisala frequency or the buoyancy frequency [2, Chap.1,2]. Here ρ(z) is the density considered as a function of the depth $z$, g is the acceleration in the gravity force field, the sign "-" originates from the increase of the density with the depth and therefore $\frac{d\rho(z)}{dz} < 0$ [3, Chap.4].

The exact solutions of the essential equations describing the internal gravity waves are only obtained for special cases [4, Chap.2-4, 5-9]. That is the reason why the approximate asymptotical methods are systematically used for the investigation of the internal gravity wave fields in stratified ocean. The internal gravity waves are usually represented in the following integral form [1, Chap.1,2][5,6] : $J = \int_\gamma \exp[\lambda f(z)] F(z) dz, \quad \lambda \gg 1$, where $f(z)$ and $F(z)$ are analytic functions of the complex variable $z$; γ is a contour of integration on the complex plane $z$. The universal way to construct the asymptotic forms of such integrals is the method of etalon integrals [10, Chap.1, 11, Chap.1,3,5].

This paper is devoted to the systematical description of a generalization of the geometrical optics method, i.e. we discuss the spatio-temporal ray method of etalon functions [10, Chap.2]. This method allows one to solve the problem of asymptotic modeling of the inharmonic wave packet's dynamics for the internal gravity waves in stratified media with slowly varying parameters. The main reasons to use the ray methods are the following: the ray representations are well correlated with the intuition and with the empirical material for the propagation of the internal gravity waves in natural stratified media (ocean, atmosphere). These methods are universal and very often one can use only them for the approximate computations of the wave fields in slowly changing non-homogeneous stratified media [10, Chap.3].

The horizontal non-homogeneity and non-stationarity are crucial for the propagation of the internal gravity waves in natural stratified media (such as the ocean and the atmosphere). To the most typical horizontal inhomogeneities of the real ocean we relate the change in the ocean bottom shape, the inhomogeneity of the density field and the variance of the mean currents. The exact solution of the problem, for example by means of separating of variables, can only be obtained when the density distribution and the ocean bottom shape are described by the simple model functions. For the arbitrary stratification and the arbitrary ocean bottom topography it is only possible to construct the asymptotic representations of the solutions [12-17].

However, if the depth or the ocean and its density vary slowly in comparison with the characteristic length (period) of the internal gravity waves, which takes place in the real ocean, then one can use the spatio-temporal ray method (the geometrical optics method) and its generalizations to investigate the mathematically modeled dynamics of the internal gravity waves. In [1, Chap.3, 8,9] the asymptotic forms of the internal gravity waves far field were obtained for the case of the constant depth. It was also shown that the far field is equal to the sum of modes, each of the modes being confined within its Mach cone. One can represent the asymptotic form of each mode via the Airy functions (the Airy wave) or the Fresnel integrals (the Fresnel wave) .

One can solve the problem using the modified spatio-temporal ray method (the geometrical optics method) proposed above. This is the method of etalon functions [10, Chap.2,3]. Its distinguishing feature is that in order to investigate the evolution of non-harmonic wave packets in stratified non-stationary horizontally non-homogeneous media one seeks for the solution in form of rational powers series with respect to the small parameter. The powers depend on the form of representation for the wave packet. The form of representation is determined by the asymptotic behavior of the solution in the stationary horizontally homogeneous case [1, Chap.3]. The phase of the wave packet can be obtained from the corresponding eikonal equation, which can be solved numerically on the characteristics (rays). The amplitude of the wave packet can be found from a conservation law along the characteristics (rays) [10,Chap.2],[8,9].

The slowness condition of the change in parameters of the medium in time and along the horizontal is crucial for applying the geometrical optics methods. The slowness is considered in comparison with the characteristic lengths and periods of internal gravity waves. However, these conditions are not sufficient for the geometrical optics methods to be valid. It is clear that for the estimates of the accuracy of the geometrical optics method one has to use the results obtained by a more precise approach than that of the spatio-temporal ray method. However because of the serious mathematical difficulties it is not yet possible. For the investigation of the dynamics of inharmonic internal gravity wave packets in stratified non-homogeneous and non-stationary media we have at hand the analytic methods which are limited and do not allow one to estimate the accuracy of the geometrical optics method for the real media. In the general case there are no exact solutions, and the known rigorous solutions just indicate a possible value of inaccuracy for typical cases. The same results for the value of inaccuracy of the spatio-temporal ray method can be obtained comparing the asymptotic results with the approximate, but more general than that of the ray method, solutions of the basic wave problems. Therefore the validity of the spatio-temporal method and of its results follows from the comparison of the results with the data of natural experiments [1, Chap.5].

Investigating the evolution of internal gravity wave packets in stratified media with slowly varying parameters it is usually supposed that the packet is locally harmonic. In contrast to the most of the works devoted to the subject the modified geometrical optics method of etalon functions elaborated here and its modifications in form of decomposition into some special functions gives an opportunity to describe the structure of the wave fields near and far from wave fronts [5,6,8].

By using one of the modifications of the geometrical optics method solves the problem of construction of the uniform asymptotic form for the far field in case of the smoothly varying bottom. This is done with the use of the asymptotic representations of the wave field for the large distances from the source in case of the constant bottom topography. We call such modification "vertical modes – horizontal rays" [10, Chap.4]. In this method one does not suppose that there is slowness in the vertical direction. The solutions are represented as an expansion in waves of the special form, the Airy waves, and describe not only the evolution of the non-harmonic wave packets when they propagate above the slowly transforming ocean bottom but also the structure of the wave field of each particular mode either close or far from the wave fronts of the modes. The argument of each Airy wave is determined by the solution of the corresponding eikonal equation. The amplitude of the wave field is obtained from the energy conservation law along the ray tube [1, Chap.3],[ 8,9].

The exact analytical expressions for the rays are obtained and the features of the phase structure for the wave field are analyzed for model forms of the stratification and the bottom shape, describing the typical structure of the ocean shelf. It is shown, in particular, that different features of the wave field structures can be displayed depending on the trajectory of the source, the ocean bottom shape and the its stratification. The "blocking" spatial effect with respect to the low frequency modes of the wave field generated by the source moving along the shore of the ocean with the overcritical velocity is analyzed.

## 2. Main equations

Let us consider a non-viscous incompressible non-homogeneous liquid. If it is unperturbed we denote its density by $\rho_0(z)$ (the stratification is supposed to be stable, i.e. $\rho_0'(z) < 0$, the axis $z$ is directed downward from the liquid's surface).

The system of the hydrodynamic equations takes the following form [2,Chap.1],[3,Chap.4]

$$\frac{\partial \mathbf{v}}{\partial t} + (\mathbf{v} \cdot \nabla)\mathbf{v} = -\frac{\nabla p}{\rho} + \mathbf{g}$$

$$div\, \mathbf{v} = 0$$

$$\frac{\partial \rho}{\partial t} + \nabla \rho \cdot \mathbf{v} = 0$$

The linearized system can be expressed as follows [1, Chap.1],[4, Chap.2]

$$u_t' + \frac{1}{\rho_0(z)} p_x' = 0$$

$$v_t' + \frac{1}{\rho_0(z)} p_y' = 0$$

$$w_t' + \frac{1}{\rho_0(z)} p_z' + g\frac{\rho}{\rho_0(z)} = 0 \quad (2.1)$$

$$u_x' + v_y' + w_z' = 0$$

$$\rho_t' + \rho_{0_z}' \cdot w = 0$$

where $\mathbf{v} = (u, v, w)$ is the velocities' vector, $\mathbf{g} = (0, 0, g)$ is the gravitation acceleration vector, $p$ and $\rho$ are the deviations of the pressure and the density from their equilibrium values [2,Chap.1],[ 7,8]. A general analysis of the system (2.1) allows to obtain the equation for the vertical component of velocity [1,Chap.1],[ 3,Chap.4]

$$\Delta w_{tt} + w_{zztt} - \frac{N^2(z)}{g} w_{ztt} + N^2(z)\Delta w = 0 \quad (2.2)$$

$$\Delta = \frac{\partial^2}{\partial x^2} + \frac{\partial^2}{\partial y^2}, \quad N^2(z) = -\frac{g}{\rho_0(z)} \rho_0'(z)$$

where $N^2(z)$ - Brunt-Vaisala frequency. Let us estimate the third member from (2.2) in comparison with the second one. If we introduce the characteristic vertical scale $L_z$, then the ratio of the third member to the second one is

$$\delta = \frac{N^2 L_z}{g}$$

If $\delta \ll 1$ (in the real ocean one has $\delta \sim 10^{-2}$), then (2.2) becomes

$$\Delta w_{tt} + w_{zztt} + N^2(z)\Delta w = 0 \qquad (2.3)$$

and the corresponding approximation is called the Boussinesq approximation [2, Chap.1,2].

It is interesting to note that when $N^2(z) = N^2 = const$ and when the depth $H$ is constant, then

$$\delta = \frac{N^2 H^2}{gH} = \frac{\pi^2 c_{max}^2}{c^2}$$

Here $c = \sqrt{gH}$ is the velocity of the long waves on the surface of liquid with depth H [4, Chap.3], and $c_{max} = \frac{NH}{\pi}$ is the maximal group velocity of the internal waves. In the real ocean one has $c \sim 100\, m/\sec$, and $c_{max} \sim 1\, m/\sec$ [2, Chap.3],[12]. From system (2.1) it is possible to obtain that horizontal components of velocity $u$ and $v$ are coupled to the vertical component by the ratio [1,Chap.1],[8]

$$\frac{\partial}{\partial t}(\Delta v + w_{zy}) = 0, \quad \frac{\partial}{\partial t}(\Delta u + w_{zx}) = 0$$

**3. The equations of internal gravity waves for a point moving mass source**

Let us consider a layer in the liquid with Brunt-Vaisala frequency $N(\tilde{z})$. Let this layer be bounded by the surface $\tilde{z} = 0$ and the bottom $\tilde{z} = \tilde{H}(\tilde{x}, \tilde{y})$. Let the point mass source with intensity Q (here Q is the volume discharge per second) moves at depth $\tilde{z}_0$ uniformly and rectilinear with the velocity $V$ in the negative direction of abscissa.

Then the velocity field in the Boussinesq approximation satisfies the following system of equations [1, Chap.1]

$$\frac{\partial^2}{\partial t^2}\left(\Delta \tilde{w} + \frac{\partial^2}{\partial \tilde{z}^2}\tilde{w}\right) + N^2(\tilde{z})\Delta\tilde{w} = Q\delta''_{tt}(\tilde{x}+Vt)\delta(\tilde{y}-\tilde{y}_0)\delta'(\tilde{z}-\tilde{z}_0)$$

$$\Delta\tilde{u} + \frac{\partial^2 \tilde{w}}{\partial \tilde{\xi}\,\partial \tilde{z}} = Q\delta'(\tilde{\xi})\delta(\tilde{y}-\tilde{y}_0)\delta(\tilde{z}-\tilde{z}_0)$$

$$\Delta\tilde{v} + \frac{\partial^2 w}{\partial \tilde{y}\,\partial \tilde{z}} = Q\delta(\tilde{\xi})\delta'(\tilde{y}-\tilde{y}_0)\delta(\tilde{z}-\tilde{z}_0)$$

(3.1)

$$\Delta = \frac{\partial^2}{\partial \tilde{\xi}^2} + \frac{\partial^2}{\partial \tilde{y}^2}, \quad \tilde{\xi} = \tilde{x} + Vt$$

At the layer's boundaries the following conditions should be satisfied [13-16]

$$\tilde{w} = 0, \text{ at } \tilde{z} = 0$$

$$\tilde{w} = \tilde{u}\frac{\partial \tilde{u}}{\partial \tilde{x}} + \tilde{v}\frac{\partial \tilde{H}}{\partial \tilde{y}}, \text{ at } \tilde{z} = \tilde{H}(\tilde{x}, \tilde{y}) \quad (3.2)$$

Further we consider the case when $N = \text{const}$, $\tilde{H}(\tilde{y}) = \beta \tilde{y}$ (i.e. when the depth depends linearly on one coordinate $\tilde{y}$, and the source moves along the line $\tilde{z} = \tilde{z}_0$, $\tilde{y} = \tilde{y}_0$).

Introducing the horizontal scale

$$L_y = L_\xi = \frac{V}{N}\frac{\pi}{\beta} = \frac{V}{N}\lambda \quad \left(\lambda = \frac{\pi}{\beta}, \lambda \gg 1\right)$$

and the vertical scale $L_z = \frac{V}{N}y_0$, we have the following relations of the non-dimensional coordinates and the dimensional ones

$$\xi = \tilde{\xi}\frac{N}{V\lambda}, \quad y = \tilde{y}\frac{N}{V\lambda}, \quad z = \tilde{z}\frac{N}{Vy_0}; \quad H(y) = \frac{\pi y}{y_0}$$

$$(u, v, w) = \frac{(\tilde{u}, \tilde{v}, \tilde{w})V^2}{QN^2}$$

We note that $y = \frac{1}{M}$, where M is the Mach number $M = \frac{V}{c_{\max}}$, and $c_{\max}$ is the maximal value of the group velocity for the internal gravity waves $c_{\max} = \frac{N\tilde{H}(\tilde{y})}{\pi} = Vy$ [1, Chap.1,2].

Thus, the domain $y < 1$, corresponds to $M > 1$ (i.e. the source's velocity is greater than $c_{\max}$), and the domain $y > 1$ corresponds to $M < 1$ (i.e. $c_{\max}$ is greater than the velocity of the source $V$).

In the non-dimensional coordinates the system of equations (3.1) and the system of boundary conditions (3.2) take the forms

$$\frac{\partial^2}{\partial \xi^2}\left(\frac{1}{\lambda^2}\Delta + \frac{1}{y_0^2}\frac{\partial^2}{\partial z^2}\right)w + \Delta w = \frac{1}{\lambda^2 y_0^2}\delta''_{\xi\xi}(\xi)\delta(y - y_0)\delta'_z(z - z_0)$$

$$\frac{1}{\lambda}\Delta v + \frac{1}{y_0}\frac{\partial^2 w}{\partial y \partial z} = \frac{1}{\lambda^2 y_0}\delta(\xi)\delta'(y - y_0)\delta(z - z_0) \quad (3.3)$$

$$\frac{1}{\lambda}\Delta u + \frac{1}{y_0}\frac{\partial^2 w}{\partial \xi \partial z} = \frac{1}{\lambda^2 y_0}\delta'(\xi)\delta(y - y_0)\delta(z - z_0)$$

$$w = 0 \text{ at } z = 0 \quad (3.4)$$

$$w = \frac{\pi}{\lambda}v \text{ at } z = H(y)$$

Since the boundary conditions (3.4) does not depend on the horizontal velocity, the system (3.3) splits into two first equations and the separated third one. If the two first equations are solved, the function $u$ can be easily obtained from the third equation.

We seek for the solutions of the first two equations of system (3.3) in the form of the sum of wave modes [1, Chap.3],[9]

$$w = \sum_{n=1}^{\infty} w^n, \quad v = \sum_{n=1}^{\infty} v^n$$

Further we consider only the dominant (first) mode omitting upper index $n$. We seek a solution of the form [10, Chap.2]

$$w = \int F(z, y, \omega) e^{i\lambda(\omega\xi - s(y,\omega))} d\omega$$

$$v = \int \Psi(z, y, \omega) e^{i\lambda(\omega\xi - s(y,\omega))} d\omega \tag{3.5}$$

where

$$F = F_0(z, y, \omega) + \frac{i}{\lambda} F_1(z, y, \omega) + O\left(\frac{1}{\lambda^2}\right)$$

$$\Psi = i\Psi_1(z, y, \omega) + O\left(\frac{1}{\lambda}\right)$$

We emphasize that functions $w$ and $v$ depend on $\xi$ only via the phases of elementary solutions, superposition of which gives the complete solutions $F$ and $\Psi$. Therefore the problem is reduced to finding functions $F(z, y, \omega)$ and $s(y, \omega)$.

Let us substitute $w$ and $v$ from (3.5) in the first two equations of (3.3) and in the boundary conditions (3.4), then equate the members at the corresponding powers of $\lambda$ [1, Chap.3]. Following this way we obtain two equations for functions $F_0(z, y, \omega)$ and $F_1(z, y, \omega)$

$$F_{0_{zz}}'' + \frac{\omega^2 + s_y'^2}{\omega^2} \cdot (1 - \omega^2) y_0^2 F_0 = 0$$

$$F_0\big|_{z=0} = 0, \quad F_0\big|_{z=H(y)} = 0 \tag{3.6}$$

$$F_{1_{zz}}'' + \frac{\omega^2 + s_y'^2}{\omega^2} \cdot (1 - \omega^2) y_0^2 F_1 = \frac{(1 - \omega^2) y_0^2}{\omega^2} \left(2 F_{0_y}' s_y' + F_0 s_{yy}''\right)$$

$$F_1\big|_{z=0} = 0, \quad F_1\big|_{z=H(y)} = -\frac{\pi s_y'}{y_0(\omega^2 + s_y'^2)} F_{0_z}' \tag{3.7}$$

Let us consider the Sturm-Liouville problem (3.6). The quantification condition yields (we remind that the dominant mode is considered, i.e. $n = 1$) [2, Chap.4]:

The eikonal equation

$$\omega^2 + s_y'^2(y, \omega) = k^2(\omega, y),$$

The dispersion relation

$$k^2(\omega, y) = \frac{\omega^2}{(1-\omega^2)y^2},$$

and the eigenfunction $F_0(z, y, \omega) = c(y, \omega)\sin\frac{zy_0}{y}$, where $c(y, \omega)$ is yet an arbitrary function which does not depend on $z$.

## 4. Finding the factor $c(y, \omega)$ from the conservation law and the locality principle.

In order to find out how the function $c(y, \omega)$ depends on y we multiply equation (3.7) by $F_0$ and then integrate the result by z from 0 to $H(y)$. Integrating by parts the left-hand member yields

$$F'_{0z}\big|_{z=H(y)} \cdot \frac{\pi s'_y(y,\omega)}{y_0 k^2(y,\omega)} = -\frac{1-\omega^2}{\omega^2} y_0^2 \int_0^{H(y)} \frac{\partial}{\partial y}(s'_y F_0^2)\, dz \qquad (4.1)$$

where

$$\int_0^{H(y)} \frac{\partial}{\partial y}(s'_y F_0^2)\, dz = \frac{\partial}{\partial y}(s'_y \Phi),$$

$$\Phi = \int_0^{H(y)} F_0^2\, dz = C^2(y,\omega)\frac{\pi}{2}\cdot\frac{y}{y_0} \qquad (4.2)$$

In order to find $F'_{0z}\big|_{z=H(y)}$ from the left-hand member of (4.1), we differentiate equation (3.6) by y, multiply the result by $F_0$ and then integrate by parts from 0 to $H(y)$. Hence we obtain

$$y_0^2 \cdot \frac{1-\omega^2}{\omega^2}\frac{\partial}{\partial y}\left(k^2(y,\omega)\right)\cdot \Phi + F'_{0z}\big|_{z=H(y)}\cdot\frac{\partial H}{\partial y} = 0$$

From here it follows

$$\frac{\pi}{y_0}\frac{s'_y}{k^2}F'_{0z}\big|_{z=H(y)} = -y_0^2 \cdot \frac{1-\omega^2}{\omega^2}\frac{\partial k^2}{\partial y}\cdot\Phi\cdot\frac{s'_y}{k^2} \qquad (4.3)$$

Substituting (4.3) and (4.2) to (4.1) yields

$$\Phi s'_y \left(\log k^2\right)'_y = \frac{\partial}{\partial y}(s'\Phi)$$

The equation $\dfrac{s'_y \Phi}{k^2(y,\omega)} = const(\omega, y_0)$ is the conservation law [1, Chap.3].

Rewriting the conservation law as

$$\frac{s'_y(y,\omega)\dfrac{\pi}{2}\dfrac{y}{y_0}c^2(y,\omega)}{k^2(y,\omega)} = \frac{s'_y(y_0,\omega)\dfrac{\pi}{2}c^2(y_0,\omega)}{k^2(y_0,\omega)}$$

we find

$$c(y,\omega) = c(y_0,\omega)\frac{y_0}{y}\sqrt[4]{\frac{1-(1-\omega^2)y_0^2}{1-(1-\omega^2)y^2}}$$

Using the locality principle, i.e. the known solution of the problem for the constant depth $H(y_0) = \pi$ [8,9], we determine $c(y_0,\omega)$ in the form

$$c(y_0,\omega) = \frac{2i\omega \sin z_0}{\pi\sqrt{1-\omega^2}\sqrt{1-(1-\omega^2)y_0^2}}$$

Finally we have

$$F_0(y,\omega,z) = \frac{2i}{\pi}\frac{y_0}{y}\frac{\omega}{\sqrt{1-\omega^2}\sqrt[4]{1-(1-\omega^2)y^2}\sqrt[4]{1-(1-\omega^2)y_0^2}}\sin z_0 \sin\left(\frac{zy_0}{y}\right) \tag{4.4}$$

## 5. The field in the domain of the subcritical velocity $(y<1)$. Rays and the stationary phase method

Let us consider the eikonal equation

$$s'^2_y(y,\omega) = \frac{\omega^2}{(1-\omega^2)y^2} - \omega^2$$

From here we get

$$s(y,\omega) = \int_{y_0}^{y}\frac{\omega\sqrt{1-y^2\alpha}}{y\sqrt{\alpha}}dy = \tag{5.1}$$

$$= \frac{\omega}{\sqrt{\alpha}}\left(\left(\sqrt{1-y^2\alpha} - \log\left(1+\sqrt{1-y^2\alpha}\right) + \log y\right) - \left(\sqrt{1-y_0^2\alpha} - \log\left(1+\sqrt{1-y_0^2\alpha}\right) + \log y_0\right)\right)$$

$$\alpha = 1-\omega^2$$

Then the field of the incident wave has the following form (in the domain $y_0 < y < 1$)

$$w = \int_0^1 \frac{4}{\pi}\frac{y_0}{y}\frac{\omega}{\sqrt{1-\omega^2}\sqrt[4]{1-y^2(1-\omega^2)}\sqrt[4]{1-y_0^2(1-\omega^2)}}\sin z_0 \sin\frac{zy_0}{y}\cos\left(\lambda[\omega\xi - s(y,\omega)] + \frac{\pi}{2}\right)d\omega$$

$$\tag{5.2}$$

To investigate the integral (5.2) at large values of $\lambda$ by the stationary-phase method we introduce the family of rays. We define it as the set of points at which the phase $\chi(\omega,\xi,y) = \lambda(\omega\xi - s(y,\omega)) + \dfrac{\pi}{2}$ is stationary, i.e. $\dfrac{\partial \chi(\omega,\xi,y)}{\partial \omega} = 0$ or

$$\xi(y,\omega) = \dfrac{\partial s(y,\omega)}{\partial \omega}.$$

Differentiating (5.1) by $\omega$, we obtain the desired rays' family

$$\xi(y,\omega) = \dfrac{\alpha\sqrt{1-y^2\alpha} - \log\!\left(1+\sqrt{1-y^2\alpha}\right) + \log y - \alpha\sqrt{1-y_0^2\alpha} + \log\!\left(1+\sqrt{1-y_0^2\alpha}\right) - \log y_0}{\alpha^{3/2}}$$

(5.3)

The expression (5.3) determines the one-parameter family of ascending rays on the plane $\xi, y$. These rays start at the point $\xi = 0, y = y_0$ with the parameter $\lambda$ (or $\omega$). At a certain value of $\omega$ we have the certain ray.

However, the expression (5.3) describes the ray's behavior only before the turning point

$$y_* = \dfrac{1}{\sqrt{\alpha}}, \quad \xi_* = \xi(y_*,\alpha) \text{ (i.e. at } \xi \le \xi_*\text{)}.$$

The reflected ray ($\xi \le \xi_*$) is constructed as follows. First we determine the eikonal of the reflected ray $s_1(y,\omega)$

$$s_1(y,\omega) = s(y_*,\omega) - \int_{y_*}^{y} \dfrac{\omega\sqrt{1-y^2\alpha}}{y\sqrt{\alpha}} dy =$$

$$= \dfrac{\omega}{\sqrt{\alpha}}\Big(\!\left(-\sqrt{1-y^2\alpha} + \log\!\left(1+\sqrt{1-y^2\alpha}\right) - \log y - \log\alpha\right) -$$

$$- \left(\sqrt{1-y_0^2\alpha} - \log\!\left(1+\sqrt{1-y_0^2\alpha}\right) + \log y_0\right)\!\Big)$$

(5.4)

Then for the reflected field $w_1$ in the whole domain $0 < y < 1$ we obtain the following expression

$$w_1 = \int_0^1 \dfrac{4}{\pi}\dfrac{y_0}{y}\dfrac{\omega}{\sqrt{1-\omega^2}}\dfrac{\sin z_0 \sin\dfrac{zy_0}{y}}{\sqrt[4]{1-y^2(1-\omega^2)}\sqrt[4]{1-y_0^2(1-\omega^2)}}\cos(\lambda(\omega\xi - s_1(y,\omega)) + \pi)d\omega$$

(5.5)

i.e. in comparison with w instead of the eikonal of the incident wave $s(y,\omega)$ we use the eikonal of the reflected wave $s_1(y,\omega)$. The phase shift $(+\pi/2)$ due to the reflection of the ray is also taken into account. The reflected ray is constructed analogously to the incident one

$$\xi_1(y,\alpha) = \dfrac{\partial s_1}{\partial \omega}$$

$$\xi_1(y,\alpha) = \dfrac{-\alpha\sqrt{1-y^2\alpha} + \log\!\left(1+\sqrt{1-y^2\alpha}\right) - \log y}{\alpha^{3/2}} +$$

$$+ \dfrac{-\alpha\sqrt{1-y_0^2\alpha} + \log\!\left(1+\sqrt{1-y_0^2\alpha}\right) - \log y_0 - \log\alpha}{\alpha^{3/2}}$$

(5.6)

Comparing (5.3) and (5.6) shows that the incident and the reflected rays are symmetric at $y_0 < y < 1$ with respect
to the line $\xi = \xi^*$ and in particular

$$\xi\big|_{y=y_*} = \xi_1\big|_{y=y_*}, \qquad \xi_1\big|_{y=y_0} = 2\xi_1\big|_{y=y_*}$$

In fig.1 two pairs of rays starting at the point $\xi = 0$, $y_0 = 0.3$ for $\alpha = 0.65$ ($\omega = 0.6$) and $\alpha = 0.45$ ($\omega = 0.75$) are represented.

The first pair for $\alpha = 0.65$ consists of the ascending ray OAB (which in turn consists of the incident ray OA and the reflected ray AB) and the descending incident ray OC.

The second pair for $\alpha = 0.45$ consists of the ascending ray ODE and the descending incident ray OF. The descending incident rays do not reflect.

In the domain $0 < y < y_0$, the reflected field is still described by the integral from (5.5). The field of the descending incident wave is determined by the formula (5.2), where we place the sign «+» at the determined by (5.1) eikonal $s(y, \omega)$.

In fig.2 and in fig.3 we represent the rays' families for two values of $y_0$: $y_0 = 0.3$ in fig.2 and $y_0 = 0.7$ in fig.3. There are the ascending rays possessing a turning point and the descending ones without a turning point. In fig.2 the value of α changes from 0.4 to 1 (which corresponds to the change in ω from 0 to 0.77) with the step 0.1. In fig.3 the value of α changes in the same limitations, but with the step 0.06. The envelope curve for the rays (the caustic surface) is also represented in the figures. It will be discussed below.

Let us now estimate the fields represented in the integral forms. We do this by using the example of the incident wave (5.2). Applying the stationary-phase method ($\lambda \gg 1$) to the integral we get

$$w(\xi, y) = \frac{B(\omega, y, y_0, z, z_0)}{\sqrt{\lambda \dfrac{\partial \xi(y, \omega)}{\partial \omega}}} \cos\left[\lambda(\omega\xi - s(y, \omega)) + \frac{\pi}{4}\right] \tag{5.7}$$

where $B(\omega, y, y_0, z, z_0) = \dfrac{4\sqrt{2}\, y_0 \omega \sin z_0 \sin\dfrac{zy_0}{y}}{\sqrt{\pi}\, y\sqrt{1-\omega^2}\, \sqrt[4]{1-y^2(1-\omega^2)}\sqrt[4]{1-y_0^2(1-\omega^2)}}$,

and the function $\xi(y, \omega)$ is determined by (5.3).

In the expression (5.7) the frequency $\omega = \omega(\xi, y)$ is implicitly determined for the prescribed values of $\xi$ and y by a solution of the equation

$$\xi = \xi(y, \omega) \tag{5.8}$$

Thus, in order to find, e.g. the incident wave at the prescribed value of y as a function of $\xi$, we need to solve the determinative equation (5.8) with respect to ω for each value of $\xi$, then to substitute the result in (5.7).

However, there is a chance to simplify significantly the finding of the field mentioned above. Namely, considering (5.7) as a function of $\omega$ only at the prescribed value of $y$: $w(\xi(y,\omega), y)$, together with the ray's equation $\xi = \xi(y, \omega)$ determined by (5.3), we can easily find the desired functional dependence $w(\xi, y)$ in the parametric way. It is possible because all functions from the formulae mentioned above are explicit.

Similarly, using the stationary-phase method the field of the reflected wave $w_1(\xi, y)$ is constructed

$$w_1(\omega, y) = -\frac{B(\omega, y, y_0, z, z_0)}{\sqrt{\lambda \frac{\partial \xi_1(y, \omega)}{\partial \omega}}} \cos\left[\lambda(\omega \xi_1(y, \omega) - s_1(y, \omega)) + \frac{3\pi}{4}\right]$$

$$\xi = \xi_1(y, \omega)$$
(5.9)

For the incident field $w_2(\xi, y)$ formed by the descending rays we have

$$w_2(\omega, y) = -\frac{B(\omega, y, y_0, z, z_0)}{\sqrt{\lambda\left(-\frac{\partial \xi(y, \omega)}{\partial \omega}\right)}} \cos\left[\lambda(-\omega \xi(y, \omega) + s(y, \omega)) + \frac{\pi}{4}\right]$$

$$\xi = -\xi(y, \omega)$$
(5.10)

We note that the radical expression in the denominator of (5.10) is positive, since in the corresponding expression for the eikonal the upper limit of integration is less than the lower one. The amplitudes of the obtained fields at large values of $\xi$ decrease as $\xi^{-1/2}$. It can be easily proved by considering the ray's equation for e.g. the incident wave (5.3). Large values of $\xi$ correspond to the small values of α (at the prescribed value of y), then $\xi \sim \alpha^{-3/2}$, and $\xi' \sim \alpha^{-5/2}$. Substituting these estimates to the amplitude of the incident wave we obtain the desired result.

In fig.4 one can see the incident wave $w_2(\xi)$ formed by the ascending rays, the reflected wave $w_1(\xi)$ and the sum of these waves (the integral field). The incident and the reflected waves are obtained by the stationary-phase method (in the parametric form) and by the numerical integration. The values of parameters used for calculations are $\lambda = 16$, $y_0 = 0.3$, $y = 0.95$, $z_0 = 1.57$, $z = 2.32$. One can see in the figure that the frequency and the amplitude of the reflected wave are less than those of the incident wave.

This becomes clear from fig.2, since the reflected ray coming to a point of intersection of two rays starts at $y_0$ at the larger value of angle than the incident ray. Therefore its optical path length is also greater than that of the incident ray. However these frequencies and amplitudes are not too different, that is why in the figure of the integral field we clearly see wave trains (or beatings).

**6. The field in the domain of the overcritical velocities ($y > 1$). The caustic surface and the uniform approximation for the integrand**

The wave space of solutions in the domain of the overcritical velocities is bounded by the line $y = 1$ and, as one can see in fig.2,3, by the caustic surface (the envelope curve of the incident ascending rays) [10, Chap.3]. Therefore, we first describe the behavior of the caustic surface.

The caustic surface is determined by the solution of system

$$\frac{\partial \xi(y,\alpha)}{\partial \omega} = 0 \qquad (6.1)$$
$$\xi = \xi(y,\alpha)$$

Here, since $\alpha = 1 - \omega^2$, the first equation from (6.1) splits into two equations

$$\omega = 0$$
$$\frac{\partial \xi(y,\alpha)}{\partial \alpha} = 0$$

Then we obtain the envelope curve itself from the following system

$$\frac{\partial \xi(y,\alpha)}{\partial \alpha} = 0 \qquad (6.2)$$
$$\xi = \xi(y,\alpha)$$

and the "beak-like" curve formed by the incident and the reflected ray at $\omega = 0$ ($\alpha = 1$)

$$\xi = \xi(0, y) \qquad (6.3)$$
$$\xi = \xi_1(0, y)$$

where $\xi(\omega, y)$ and $\xi_1(\omega, y)$ are determined by (5.3) and (5.6).

When the value of y is close to 1, the equation for the "beak-like" curve is a semi cubic parabola with the backtracking point $\xi = \xi_0$ and $y = 1$

$$(\xi - \xi_0)^2 = \frac{8}{9}(1-y)^3 \qquad (6.4)$$

where $\xi_0 = \log\left(1 + \sqrt{1-y_0^2}\right) - \sqrt{1-y_0^2} - \log y_0$.

In fig.5 one can see the "beak" itself (see equation (6.3))(line 1) and its approximation (6.4)(line 2) at $y_0 = 0.7$. We notice that the approximation (6.4) depends only on the single parameter $y_0$ (on the horizon of the source's motion).

Let us now rewrite the first equation from (6.2) in the following form

$$\sqrt{1-y^2\alpha} = \frac{1-\alpha}{\varphi(y,\alpha)} \qquad (6.5)$$

where

$$\varphi(y,\alpha) = \frac{\alpha - 1}{\sqrt{1-y_0^2\alpha}} - 3\left(\log y - \log y_0 - \log\left(1 + \sqrt{1-y^2\alpha}\right) + \log\left(1 + \sqrt{1-y_0^2\alpha}\right)\right)$$

The right-hand member in (6.5) at $\alpha$ close to 1 tends to zero, the function $\varphi(y,\alpha)$ is bounded from below with respect to $\alpha$. This allows one to apply the perturbation method to equation (6.5). In the first approximation we obtain its solution substituting the zero for the right-hand member in (6.5): $y = \frac{1}{\sqrt{\alpha}}$, or $\alpha = \frac{1}{y^2}$, which together with $\xi = \xi(y,\alpha)$, where $\alpha = 1/y^2$, gives the curve passing through the turning points of the rays

$$\xi_n(y) = -\sqrt{y^2 - y_0^2} + y^3 \log\left(\frac{y}{y_0} + \sqrt{\left(\frac{y}{y_0}\right)^2 - 1}\right) \qquad (6.6)$$

We obtain the second approximation substituting the value $1/\sqrt{\alpha}$ for y to $\varphi(y,\alpha)$

$$y = \frac{1}{\sqrt{\alpha}}\sqrt{1-\delta_2} \qquad (6.7)$$

$$\delta_2 = \frac{(1-\alpha)^2}{\varphi^2\left(\frac{1}{\sqrt{\alpha}},\alpha\right)}$$

One can find the graph of the correction for the first approximation $\delta_2$ in fig.6. The maximal value of the correction is only 3%.

The equation (6.7) together with $\xi = \xi(y,\alpha)$, where it is needed to substitute for y its value from (6.7), gives the second approximation for the caustic surface in the parametric way. The caustic surface obtained in this approximation is plotted in fig.2, 3.

Describing the field of the vertical velocity in the domain $y > 1$ it is necessary to take into account the turning point $\omega_* = \frac{\sqrt{y^2-1}}{y}$, which is a branch point on the complex plane $\omega$. At this point the amplitude of the integrand is equal to infinity. Therefore one should split the whole interval of integration by $\omega$ from 0 to 1 into two intervals: from 0 to $\omega_*$, and from $\omega_*$ to 1. The second interval of integration corresponds to the domain to the right of the caustic surface (the real rays), the first interval of integration corresponds to the domain to the left of the caustic surface (the complex rays). The incident and the reflected fields are still determined by formulae (5.2) and (5.5), but the lower limit of integration becomes equal to $\omega_*$. In order to find the integrand on the first interval, i.e. for the field of the so-called penetrating wave, we consider the analytic continuation of the integrand of the incident wave by $\omega$ through the lower half-plane (or of the reflected wave through the upper one). In any case we obtain the following expression for the penetrating wave $w_n$

$$w_2 = \operatorname{Re}\int_0^{\omega_*} \frac{4e^{i\frac{3\pi}{4}} y_0 \omega \sin z_0 \sin\frac{zy_0}{y}}{\pi y\sqrt{1-\omega^2}\,(y^2\alpha-1)^{1/4}(1-y_0^2\alpha)^{1/4}} e^{i\lambda(\omega\xi - s_2(\omega,y))}\, d\omega, \qquad (6.8)$$

where

$$s_2(\omega, y) = \frac{\omega}{\sqrt{1-\omega^2}} \left( -i\sqrt{y^2\alpha-1} - \log\left(1 - i\sqrt{y^2\alpha-1}\right) + \log y - \sqrt{1-y_0^2\alpha} + \log\left(1 + \sqrt{1-y_0^2\alpha}\right) - \log y_0 \right)$$

Thus, the integral field consists of three components

$$w_c = w + w_1 + w_2$$

Let us write the integrand $f_c$

$f_c = f + f_1$, for $\omega > \omega_*$
$f_c = f_2$, for $\omega < \omega_*$

$$f = \operatorname{Re} \frac{b(\omega, y)}{(1 - y^2\alpha)^{1/4}} e^{i\left(\lambda(\omega\xi - s(\omega,y)) + \frac{\pi}{2}\right)}$$

$$f_1 = \operatorname{Re} \frac{b(\omega, y)}{(1 - y^2\alpha)^{1/4}} e^{i(\lambda(\omega\xi - s_1(\omega,y)) + \pi)} \tag{6.9}$$

$$f_2 = \operatorname{Re} \frac{b(\omega, y)}{(y^2\alpha - 1)^{1/4}} e^{i\left(\lambda(\omega\xi - s_2(\omega,y)) + \frac{3\pi}{4}\right)},$$

$$b(\omega, y) = \frac{4 y_0 \omega \sin z_0 \sin \frac{z y_0}{y}}{\pi y \sqrt{1-\omega^2}(1 - y_0^2\alpha)^{1/4}}$$

The functions $f$, $f_1$ and $f_2$ are the elementary solutions of internal gravity equation (3.3) at the prescribed value of frequency $\omega$ in WKB approximation (geometrical optics) or, what is the same, the elementary WKB solutions.

Our goal is to find such a function $G(\omega, \xi, y)$ that it is regular at the point $\omega = \omega_*$ and its asymptotic form for large $\lambda$ coincides with the asymptotic form of the function $f_2$ to the left from the turning point and with $f + f_1$ to the right of the turning point.

In order to do so we introduce the following function

$$\Delta(\omega, y) = S(y_n, \omega) - S(y, \omega) \tag{6.10}$$

$$\Delta(\omega, y) = \frac{\omega}{\sqrt{\alpha}} \left( \log \frac{1}{\sqrt{\alpha}} - \sqrt{1 - y^2\alpha} + \log\left(1 + \sqrt{1 - y^2\alpha}\right) - \log y \right)$$

which represents the phase incursion from the point $y$ to the turning point $y_n = \frac{1}{\sqrt{\alpha}}$. We expand $\Delta(\omega, y)$ in the series with respect to the argument $\Omega = \alpha y^2$, supposing that $\Omega$ is close to 1, and take only the dominant term of the series

$$\tilde{\Delta} = \frac{\omega}{\sqrt{\alpha}} \frac{\sqrt{1 - y^2\alpha}^3}{3} \tag{6.11}$$

We rewrite the expression (6.9), factoring out the common factor

$$i\varphi(\omega,\xi) = i\left(\lambda(\omega\xi - s(y_n,\omega)) + \frac{3\pi}{4}\right)$$

$$f = b(\omega,y)\, e^{i\varphi(\omega,\xi)}\, \frac{e^{i\left(\lambda\tilde{\Delta}-\frac{\pi}{4}\right)}}{\sqrt[4]{1-y^2\alpha}}$$

$$f_1 = b(\omega,y)\, e^{i\varphi(\omega,\xi)}\, \frac{e^{-i\left(\lambda\tilde{\Delta}-\frac{\pi}{4}\right)}}{\sqrt[4]{1-y^2\alpha}} \qquad (6.12)$$

$$f_2 = b(\omega,y)\, e^{i\varphi(\omega,\xi)}\, \frac{e^{-\lambda|\tilde{\Delta}|}}{\sqrt[4]{y^2\alpha-1}}$$

It is easy to verify that the formulae (6.12) and (6.11) represent the asymptotic form of the following function

$$\tilde{G}(\omega,\xi,y) = 2\sqrt{\pi}\cos(\varphi(\omega,\xi)) b(\omega,y)\, q^{1/6}\, Ai\left(q^{2/3}(1-y^2\alpha)\right) \qquad (6.13)$$

$$q = \frac{\lambda\omega}{2\sqrt{1-\omega^2}},$$

$$Ai(x) = \frac{1}{2\pi}\int_{-\infty}^{\infty} e^{i\left(\frac{t^2}{3}-xt\right)} dt$$

where $Ai(x)$ - is the Airy function [18, Chap.9].

The function $\tilde{G}$ is the local asymptotic form, i.e. it describes the solution near the turning point $\omega = \omega_*$. It is easy to verify that the expression (6.12) is the asymptotic form of the function $\tilde{G}(\omega,\xi,y)$. Therefore it will suffice to consider the asymptotic form of the Airy function [18,Chap.9]

$$Ai(x) \sim \frac{1}{\sqrt{\pi}x^{1/4}}\cos\left(\frac{2}{3}x^{3/2}-\frac{\pi}{4}\right), \quad \text{when } x \to +\infty$$

$$Ai(x) \sim \frac{1}{2\sqrt{\pi}|x|^{1/4}}\, e^{-\frac{2}{3}|x|^{3/2}}, \quad \text{when } x \to -\infty$$

We notice that the asymptotic form (6.13) also holds in the domain of the subcritical velocities ($y<1$), when there are no turning points in the domain of integration.

In fig.7 we represent the function $\tilde{G}(\omega,\xi,y)$ ( line 2) and the WKB-solution (6.9) $f + f_1$ (line 1) to the right of the turning point, and $f_2$ to the left of it (line 1), in the domain $y > 1$ ($y = 1.15$). In fig.8 the function $\tilde{G}(\omega,\xi,y)$ (line 2) and the WKB-solution (only $f + f_1$) (line 1) in the domain $y < 1$ ($y = 0.8$) are represented. In both cases $\xi = 2$.

The uniform asymptotic form of $\tilde{G}(\omega,\xi,y)$ has the following form

$$\widetilde{G}(\omega,\xi,y) = \frac{2\sqrt{\pi}\, b(\omega,y)\cos(\varphi(\omega,y))\left(\frac{3}{2}\Delta(\omega,y)\right)^{1/6}}{\sqrt{1-y^2\alpha}} Ai\left(\left(\frac{3}{2}\Delta(\omega,y)\right)^{2/3}\right) \quad (6.14)$$

where $\Delta(\omega,y)$ is determined by (6.10).

We emphasize that the uniform asymptotic form (6.14) at small values of arguments of the Airy function is transformed to the local asymptotic form (6.13), and at large values of arguments it is transformed to the WKB expansion (6.9) [11, Chap.5]. The uniform asymptotic form also holds in the domain of subcritical velocities ($y < 1$). It is clear that when $y < y_0$ the incident (descending) wave is to be found from the other formulae.

In fig.9 we represent the uniform asymptotic form (line 2) and the WKB solution (line 1) at the prescribed $y = 1.15$ ("section" by $\omega$), and in fig.10 we represent the uniform asymptotic form (line 2) and the WKB solution (line 1) at the prescribed $\omega = 0.4$ ("section" by $y$). In fig.11 the uniform asymptotic form (line 2) and the WKB solution (line 1) are represented at $y = 0.85$ (the domain of subcritical velocities), i.e. in the case of absence of turning points. In all figures $\xi = 2$. In all cases one can clearly see the excellent coincidence of the uniform asymptotic form and the WKB approximations.

The uniform asymptotic form describes the integrand of solution $w$ in the domain $y > y_0$ at any values of $\omega$, except in the vicinity of the point $\omega = 0$, $y = 1$ (on the plane $\xi, y$ this is the vicinity where the "beak" sharpens). The argument of the Airy function is not an analytic function at the small values of $\omega$ (it behaves like $\omega^{2/3}$), and the amplitude behaves like $\omega^{7/6}$. Let us determine the behavior of the integrand at $y = 1$ and at any values of $\omega$. One can find the acceptable function which coincides with the uniform Airy function at large values of $\omega$ and which is an analytic function at small values of $\omega$, in the following form

$$F(\omega) = b(\omega,1)\frac{\Delta^{1/8}(\omega,1)D\left(\frac{1}{4},\Delta^{1/4}(\omega,1)\right)}{\sqrt{\omega}} \quad (6.15)$$

$$D(v,x) = \frac{1}{\pi}\text{Re}\, e^{-ix^4}\int_0^\infty \frac{e^{i(t-x^2)^2}}{t^v}dt$$

$D(v,x)$ is related to the function of the parabolic cylinder [18, Chap.8]. In fig.12 one can find the integrand (without the amplitude $b(\omega,y)$), which is the uniform Airy function (line 1) and determined by the formula (6.15) (line 2). It is clear that the graphs differ only in a very narrow region adjacent to $\omega = 0$.

In fig.13 we represent the same graphs but with the amplitude factors $b(\omega,y)$ and $b(\omega)$. It is clear that these graphs almost coincide. Hence, the uniform Airy function can be further applied to the integration by $\omega$. The integral field is determined by integration of the uniform Airy function with respect to $\omega$. In fig.14 the integral field $w$ obtained by integration of the WKB approximation is represented. In fig.15 one can find the field $w$ obtained by the integration of the uniform Airy asymptotic form. We see the complete coincidence of phases and the substantial divergence of the amplitudes. This results from the fact that, as it was mentioned above, the WKB solution is equal to infinity at the turning point. Despite the integrability of this singularity, the integral of the Airy function over the infinite interval and the integral of its asymptotic form over the same interval differ by approximately 23%. The wave propagating forwards in fig.14 is not valid from the physical point of view (it is absent in fig.15). This wave emerges due to the contribution of the border set of the integration domain (the turning point) to the integral. In fig.16 we see the field $w$ obtained by the integration of the WKB approximation. In fig.17 we represent the field $w$ obtained by integration of the uniform Airy asymptotic form.

It follows from the numerical results thus presented that, outside the caustic, the wave field is sufficiently small indeed and is not subjected to great many oscillations, whereas the wave picture inside the zone of caustic is a rather complicated system of incident and reflected harmonics

In Fig.18 we present the complete wave field $W$ of the vertical component of the velocity in the dimensional coordinates that is obtained by integrating the Airy uniform asymptotic (6.14) along the caustic ( the continuous line). The main parameters of the computation, which are typical for a real oceanic shelf, where chosen as follows: the slope of the bottom $\beta = 0.1$, $V = 2 m/s$, $y_0 = 500$ meters. The dashed line in this figure represents the law of diminishing amplitudes of the field on the caustic: $const\, \xi^{-10/9}$, which is obtained from the analysis, using the perturbation method, of the behavior of the integrand in (3.5). In our case, we have $const = 0.26$ and, as the numerical computations presented in this way show, the amplitude of the wave field is qualitatively correctly described by this dependence indeed.

Consider further the function

$$M(\omega, y) = \frac{\sqrt{y}\left(\frac{3}{2}\Delta(\omega,y)\right)^{1/6}}{\sqrt[4]{1-y^2(1-\omega^2)}} Ai\left(\left(\frac{3}{2}\Delta(\omega,y)\right)^{2/3}\right)$$

which coincides (up to amplitude factor) with the function $G(\omega, \xi, y)$ (6.14) and is a uniform asymptotic, with regard to the turning point, for the following generating equation:

$$\frac{\partial^2 m(y,\omega)}{\partial y^2} + \lambda^2 \omega^2 \left(\frac{1}{(1-\omega^2)y^2} - 1\right)m(y,\omega) = 0 \qquad (6.16)$$

which is regarded as an equation with a single turning point: $y_* = \frac{1}{\sqrt{1-\omega^2}}$ (in the domain $y > 0$). Equation (3.5.5.16) is obtained from the first equation of system (3.3) if we seek a solution in the form

$$W = c(y,\omega)m(y,\omega)\exp(i\lambda\omega\xi)\sin(\frac{z\,y_0}{y}).$$

We must into account here that

$$\frac{\partial m}{\partial y} \sim O(\lambda),\ \frac{\partial^2 m}{\partial y^2} \sim O(\lambda^2),\ \frac{\partial c}{\partial y} \sim O(1).$$

In particular, equation (6.16) implies the eikonal equation in the WKBJ approximation. However, the asymptotic $M(\omega, y)$ does not fit for small values of $\omega << \frac{1}{\lambda}$. To find an asymptotic behavior available for small values of $\omega$ consider the exact solution of equation (.6.16). The linearly independent solutions of the equation, $m_1$ and $m_2$ can be expressed in terms of the modified Bessel functions $I_\nu$, $K_\nu$:

$$m_1 = \sqrt{y}\, I_\nu(\lambda\omega y), \quad m_2 = \sqrt{y}\, K_\nu(\lambda\omega y)$$

with the subscript $\nu(\omega) = \dfrac{1}{2}\sqrt{\dfrac{1-(1+4\lambda^2)\omega^2}{1-\omega^2}}$, which can take both real and imaginary values, in dependence $\omega$. The Macdonald function:

$$m(y,\omega) = A(\omega)\sqrt{y}\, K_\nu(\lambda\omega y),$$

is an appropriate solution, because it decays for large positive values of $y$. To find the normalizing factor, one must consider the coincidence of the asymptotics of the functions $M(\omega,y)$ and $m(\omega,y)$ as $y \to +\infty$, namely,

$$M(\omega,y) \sim \frac{\exp(-\lambda\omega y + \lambda\omega\pi/2\sqrt{1-\omega^2})}{2\sqrt{\pi}\sqrt[4]{1-\omega^2}}$$

$$m(\omega,y) \sim A(\omega)\frac{\sqrt{\pi}\exp(-\lambda\omega y)}{\sqrt{2\lambda\omega}}$$

Using these relations one can prove that

$$A(\omega,y) = \frac{\exp(\lambda\omega\pi/2\sqrt{1-\omega^2})}{\pi\sqrt{2}\sqrt[4]{1-\omega^2}}$$

With regard to the amplitude factor obtained from the corresponding conservation law along the characteristics (rays) and from the locality principle, one can finally obtain the following asymptotic expression describing the integrand in (3.5.2.5) for any values of $\omega$:

$$M_k(y,\omega) = 2\sqrt{\pi}\, A(\omega) b(\omega,y) \cos(\varphi(\omega,\xi)) K_\nu(\lambda\omega y) \tag{6.17}$$

Obviously, the completer field $W$ of a separate mode of the vertical velocity for internal gravity waves is obtained, in accordance with (6.5), by integrating expression (6.17) with respect to $\omega$ from 0 to 1.
    Thus, in the adiabatic approximation, we have obtained computation formulas for the field of vertical velocity of the separate mode (the first one) in two zones, namely, in the zone of subcritical velocities($y<1$, first zone) and in the zone of supercritical velocities($y>1$, second zone).    In the first zone, the field is represented in the form of two integrals which are a superposition of quasiplane waves. The first of them defines the incident wave, and the other represents the reflected wave. The integrals thus obtained are evaluated by the method of stationary phase. Formulas (3.5.5.14) and (3.5.5.17) are the results of these computations. In the other zone, which includes the caustic, the field is represented by a single integral with respect to the frequency ω. The uniform asymptotic of the integrand is expressed in terms of the Airy function or the Macdonald function of a variable index depending on the frequency $\omega$ (formulas (6.14) and (6.17). The asymptotics thus obtained are analyzed numerically

## 7. Discussions

The asymptotical representations constructed above allow one to describe the far field of the internal gravity waves generated by a source moving over the slowly varying bottom. The obtained asymptotical expressions for the solution are uniform and reproduce the essential features of wave fields near caustic surfaces and wave fronts. In the paper the problem of reconstructing non-harmonic wave packets of the internal gravity waves generated by a source moving in a horizontally stratified medium is considered. The solution is proposed in terms of modes, propagating independently at the adiabatic approximation, and described as a non-integral degree series of a small parameter characterizing the stratified medium. In this study we analyze the evolution of non-harmonic wave packets of internal gravity waves generated by a moving source under the assumption that the parameters of a vertically stratified medium (e.g. an ocean) vary slowly in the horizontal direction, as compared to the characteristic length of the density. A specific form of the wave packets, which can be parameterized in terms of model functions, e.g. Airy functions, depends on a local behavior of the dispersion curves of individual modes in the vicinity of corresponding critical points.

In this paper the modified space-time ray method was proposed, which belongs to the class of geometrical optics methods. The key point of the proposed technique is the possibility to derive the asymptotic representation of the solution in terms of a non-integral degree series of the small parameter $\varepsilon = \Lambda/L$, where $\Lambda$ is the characteristic wave length, and L is the characteristic scale of the horizontal heterogeneity. The explicit form of the asymptotic solution was determined based on the principles of locality and asymptotic behavior of the solution in case of a stationary and horizontally homogeneous medium. The wave packet amplitudes are determined from the energy conservation laws along the characteristic curves. A typical assumption made in studies on the internal wave evolution in stratified media is that the wave packet is locally harmonic. A modification of the geometrical optics method, based on an expansion of the solution by model functions, allows us to describe the wave field structure both far from and at the vicinity of the wave front.

Using the asymptotic representation of the wave field at a large distance from a source moving in a layer of constant depth, we solve the problem of constructing the uniform asymptotics of the internal waves in a medium of varying depth. The solution is obtained by modifying the previously proposed "vertical modes-horizontal rays" method, which avoids the assumption that the medium parameters vary slowly in a vertical direction. The solution is parameterized through the Airy waves. This allows to describe not only the evolution of the non-harmonic wave packets propagating over a slow-varying fluid bottom, but also specify the wave field structure associated with an individual mode both far from and close to the wave front of the mode. The Airy function argument is determined by solving the corresponding eikonal equations and finding vertical spectra of the internal gravity waves. The wave field amplitude is determined using the energy conservation law, or another adiabatic invariant, characterizing wave propagation along the characteristic curves.

Modeling typical shapes and stratification of the ocean shelf, we obtain the analytic expressions describing the characteristic curves and examine characteristic properties of the wave field phase structure. As a result it is possible to observe some peculiarities in the wave field structure, depending on the shape of ocean floor, water stratification and the trajectory of a moving source. In particular, we analyze a spatial blocking effect of the low-frequency components of the wave field, generated by a source moving alongshore with a supercritical velocity. Numerical analyses that are performed using typical ocean parameters reveal that actual dynamics of the internal gravity waves are strongly influenced by horizontal inhomogeneity of ocean bottom. In paper we used an analytical approach, which avoids the numerical calculation widely used in analysis of internal gravity wave dynamics in stratified ocean.

In this paper we investigate the problems of wave theory with the help of geometrical optics methods (WKB method) and its modification. The main question consists in finding of asymptotic solution near special curve (or surface), which is called caustic. It is well known, that a caustic is an envelope of a family of rays, and asymptotic solution is obtained along this rays. Each point of the caustic corresponds to a specified ray, and that ray is tangent at this point.

It is general rule that caustic of a family of rays single out area in space, so that rays of that family can't appear in the marked area. There is also another area, and each point of this area has two rays, that pass through this point. One of those rays has already passed this point, and another is going to pass the point. Formal approximation of geometrical optics or WKB approximation can't be applied near caustic, because rays merge together in that area, after they were reflected by the caustic. If we want to find a wave field near the caustic, then it is necessary to use a special approximation of the solution, and in the paper modified ray method was proposed to build an uniform asymptotic expansion of integral forms of the internal gravity wave field.

After the rays were reflected by the caustic, there is a leap of phase. It is clear, that a leap of phase can only happen in the area, where methods of geometrical optics, which were used in previous sections, can't be applied. If rays touch caustic several times, then additional phase shifts will be summarized. Phase shift, which was created by caustic, is rather small in comparison with the change in phase along the ray. But this shift can considerably affect the interference pattern of the wave field.

The results of this paper represent a significant interest for physics and mathematics. Besides that interest asymptotic solutions, which were obtained in this paper, can present significant importance for engineering applications, since method of geometrical optics which were modified to calculate the wave field near caustic, make possible the description of different wave fields in the rather big class of another problems.

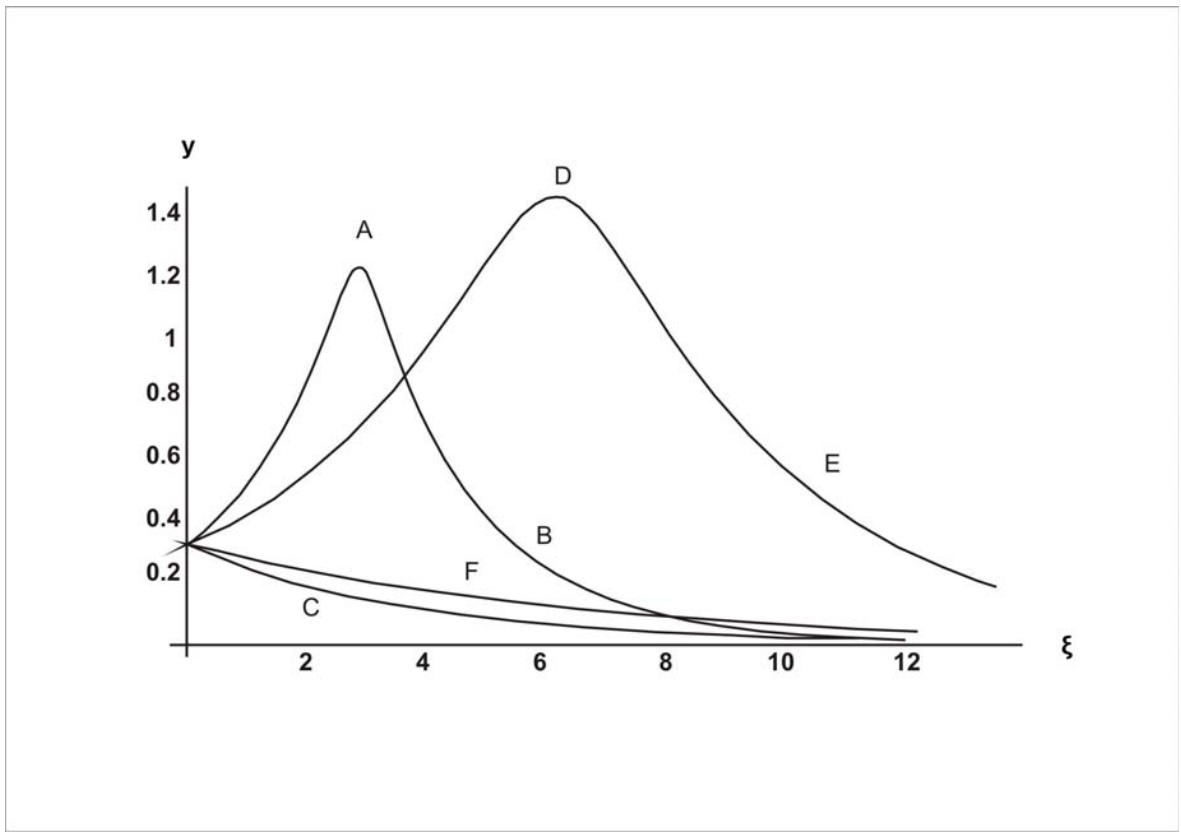

Fig.1. Rays from a moving source for and $y_0 = 0.3$ and $\omega = 0.6$ (line AB , C) , $\omega = 0.75$ (line DE, F).

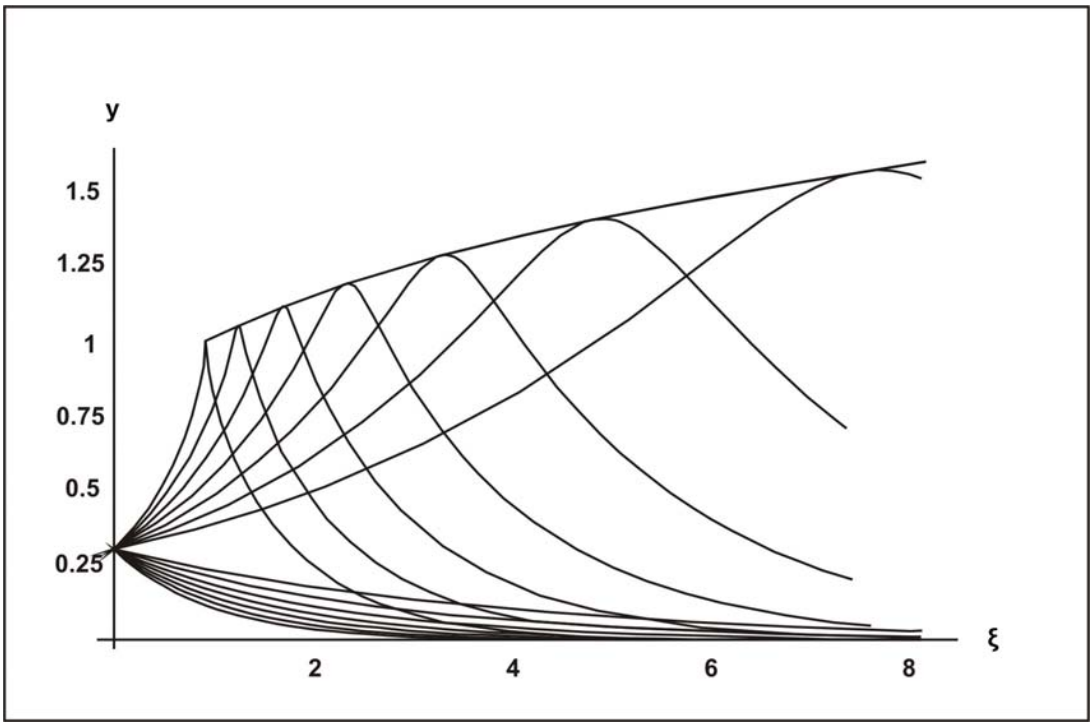

Fig.2 Rays from a moving source for $y_0 = 0.3$, value of $\omega$ change from $0$ to $0.77$ with step $0.11$.

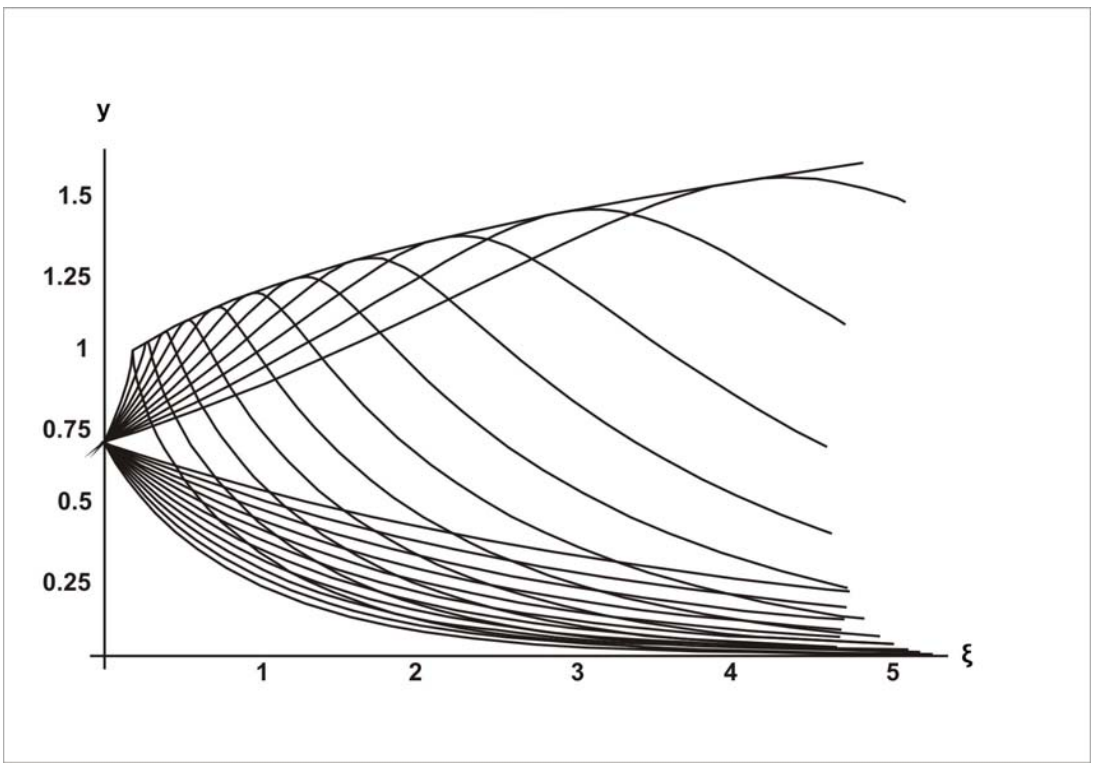

Fig.3 Rays from a moving source for $y_0 = 0.6$, the value of $\omega$ change from $0$ to $0.77$ with step $0.07$.

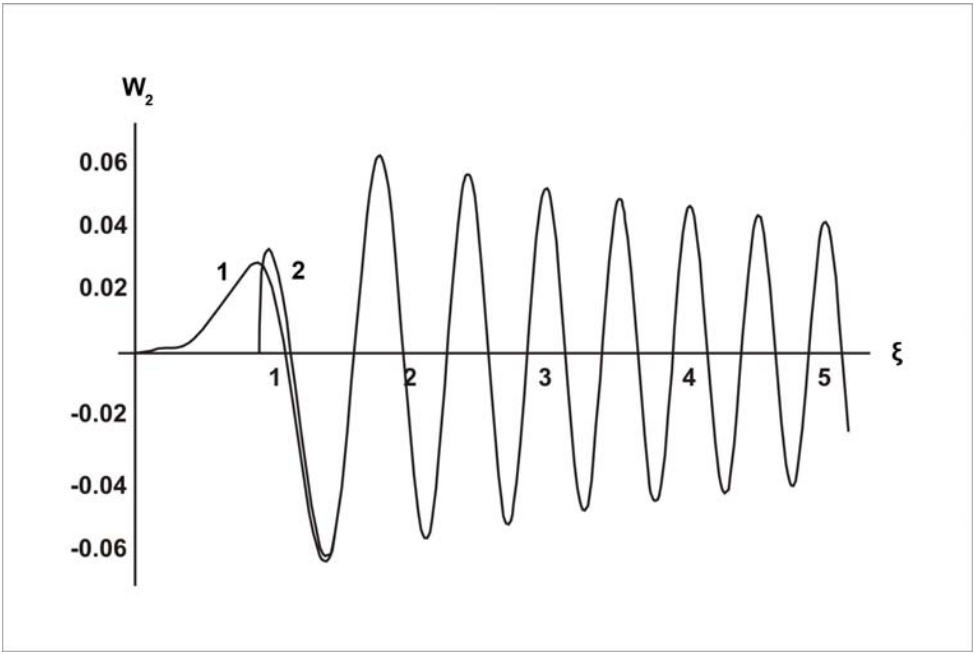

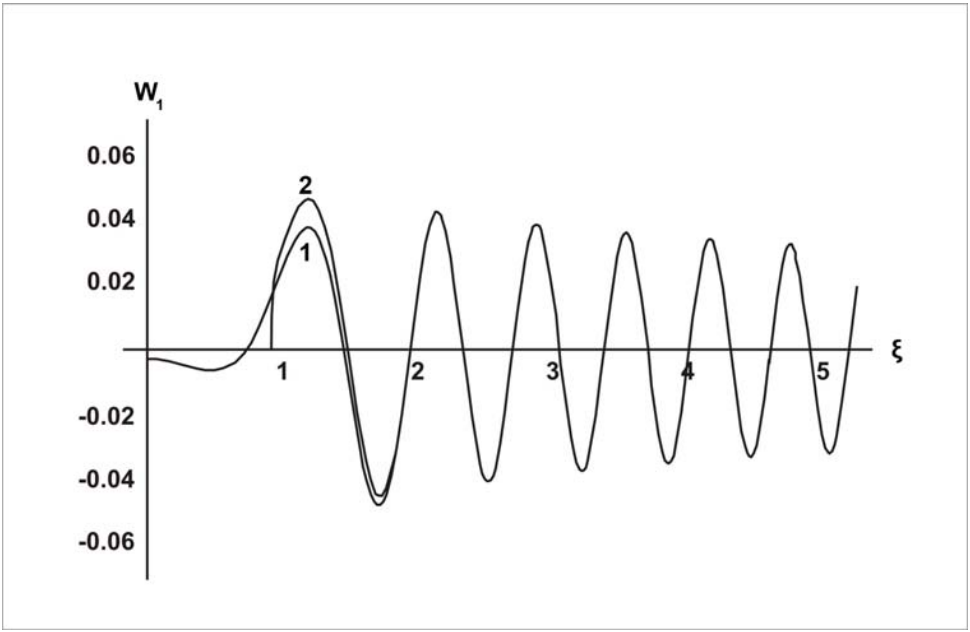

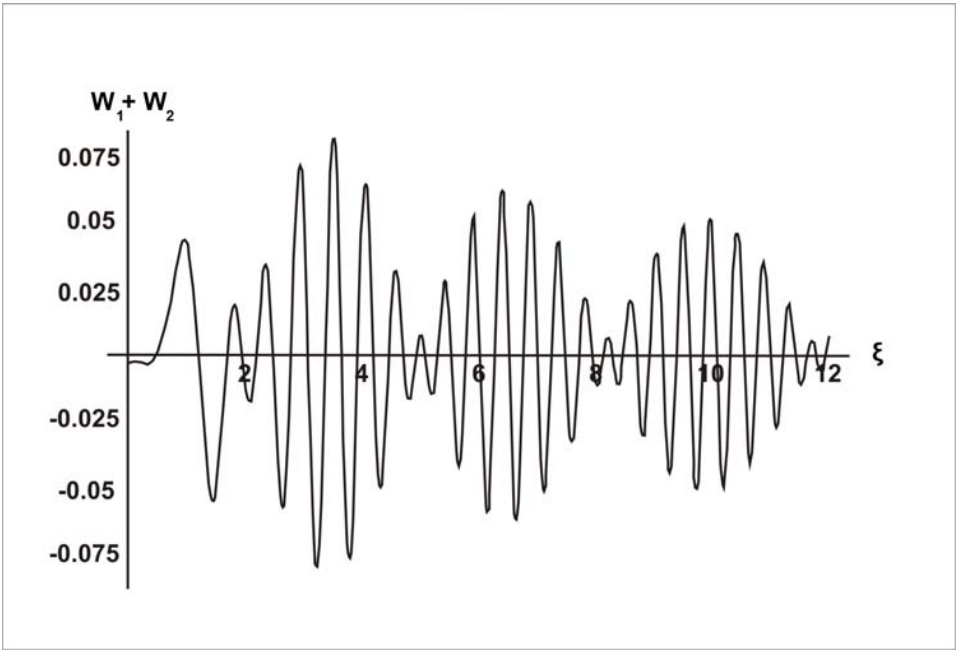

Fig.4. Vertical component $w$ of the velocity along $\xi$ axis.

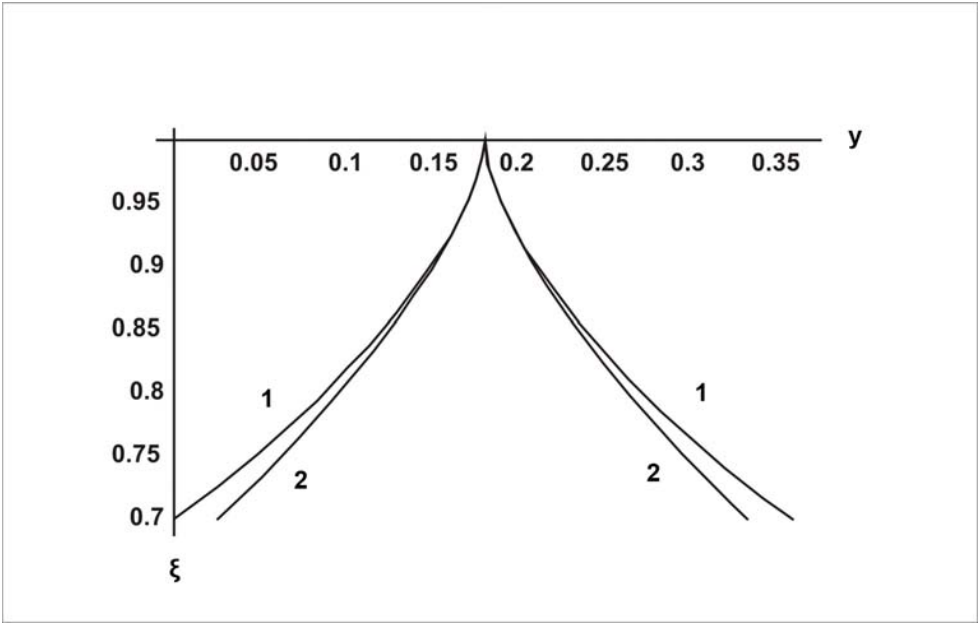

Fig.5. Caustic curve and its approximation

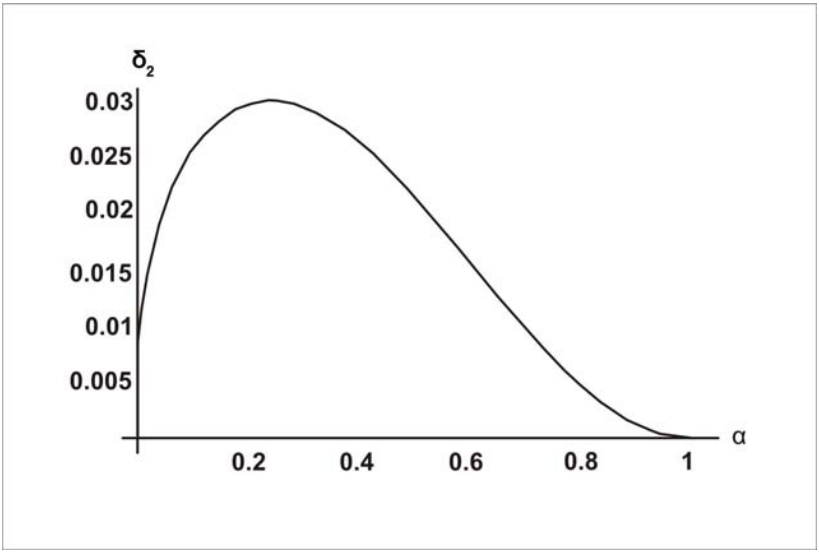

Fig.6. Correction to the caustic curve

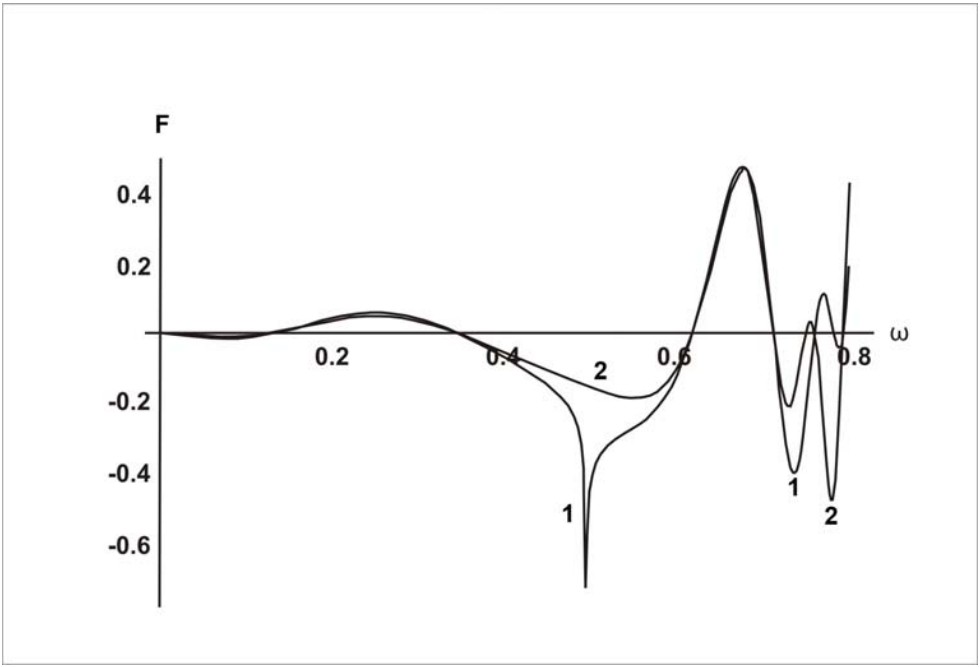

Fig.7 The local asymptotic and the WKB approximation of the vertical component velocity $w$ integrand for $y > 1$.

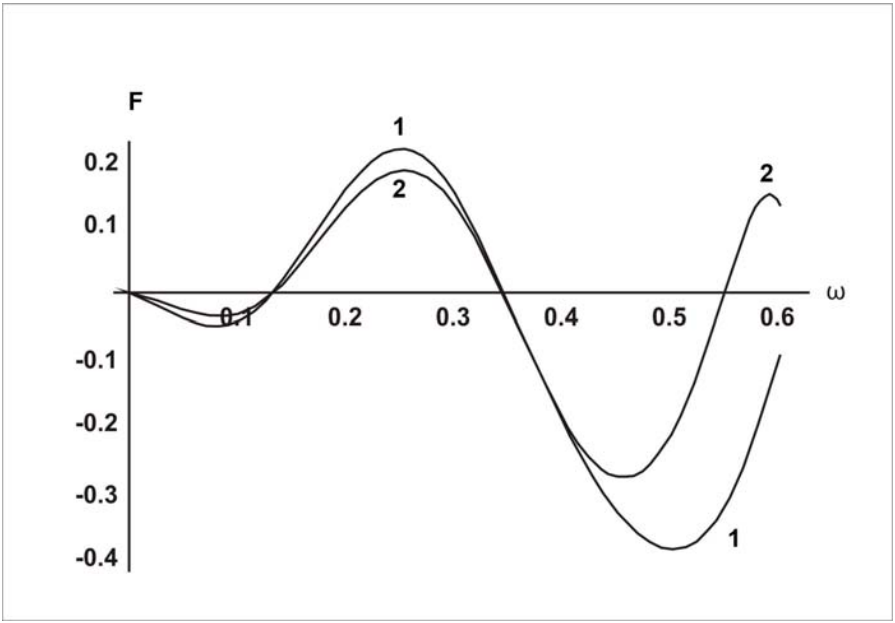

Fig.8 The local asymptotic and the WKB approximation of the vertical component velocity $w$ integrand for $y < 1$.

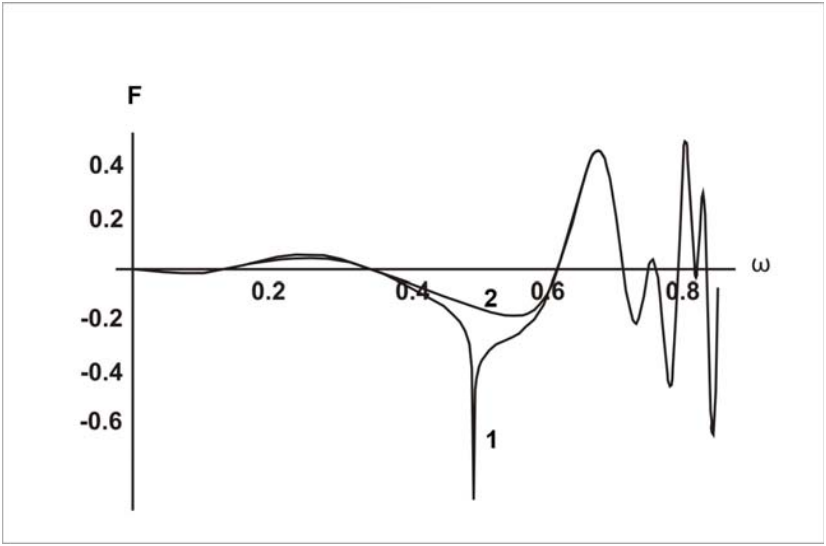

Fig.9 The uniform asymptotic and the WKB approximation of the vertical component velocity $w$ integrand along $\omega$ axis and constant $y$ in domain of critical velocities.

.

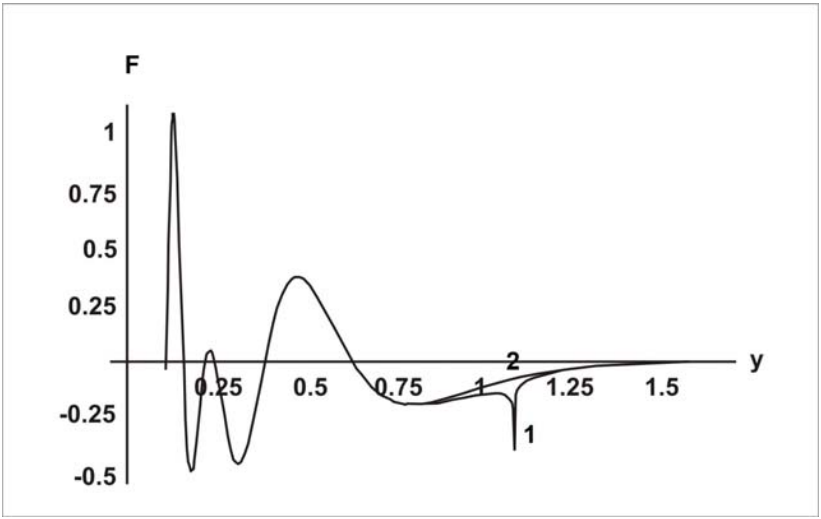

Fig.10 The uniform asymptotic and the WKB approximation of the vertical component velocity $w$ integrand along $y$ axis and constant $\omega$.

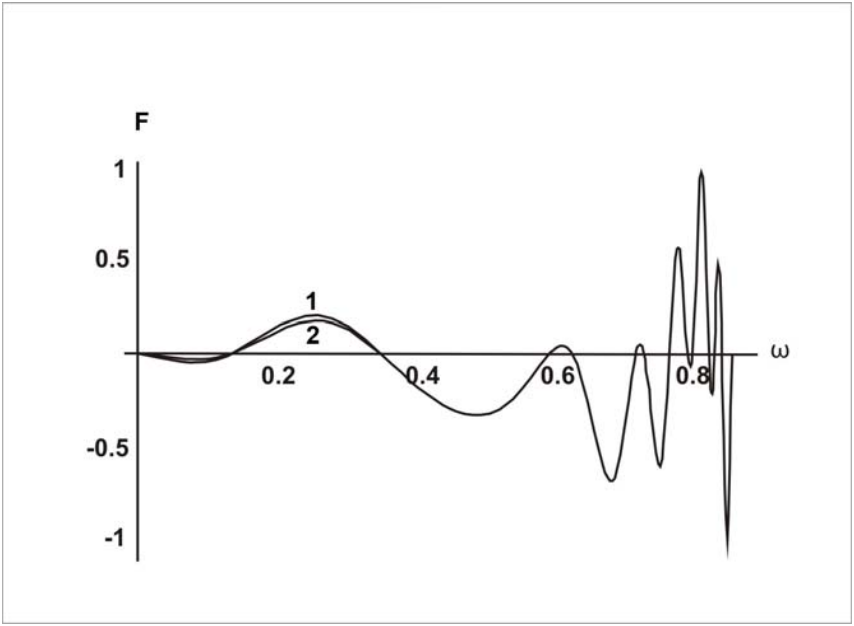

Fig.11 The uniform asymptotic and the WKB approximation of the vertical component velocity $w$ integrand along $\omega$ axis and constant $y$ in domain of subcritical velocities.

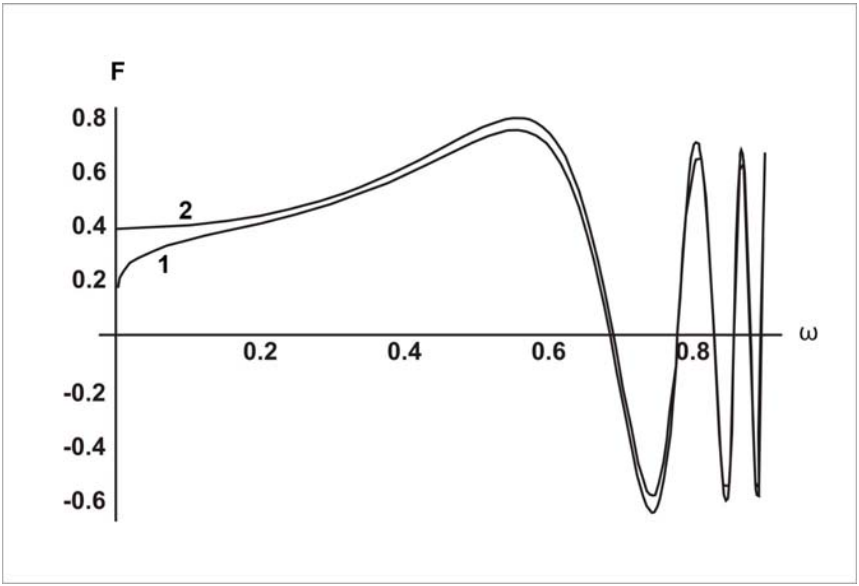

Fig.12 The uniform Airy and parabolic cylinder approximations of the vertical component velocity $w$ integrand without the amplitude factors.

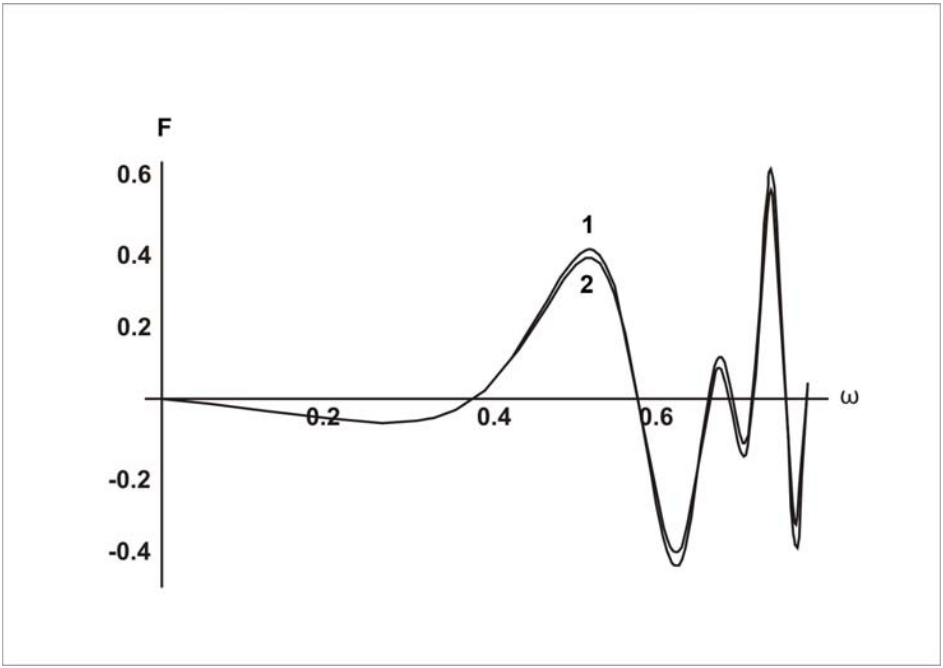

Fig.13 The uniform Airy and parabolic cylinder approximations of the vertical component velocity $w$ integrand with the amplitude factors.

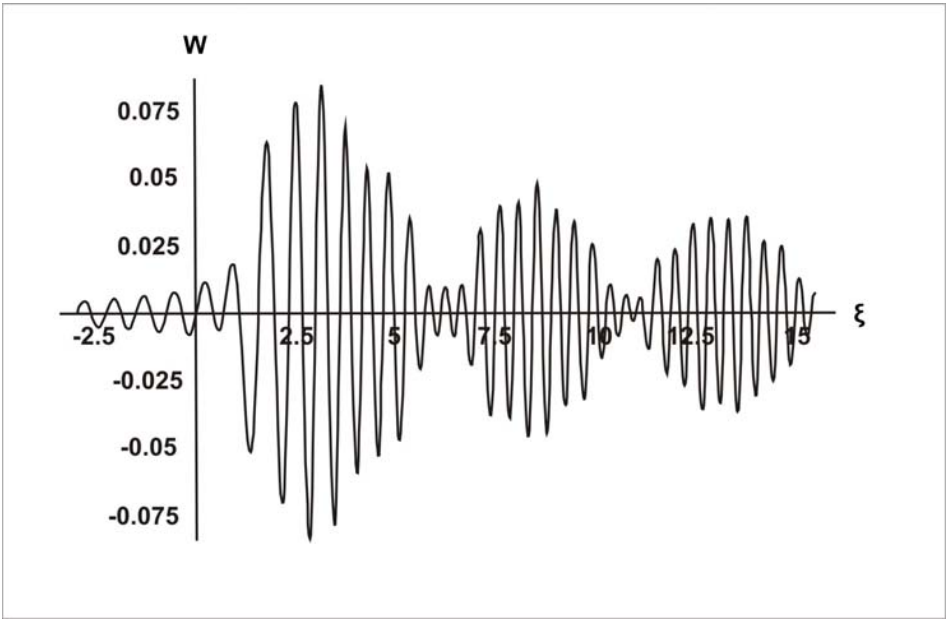

Fig.14 Vertical component $w$ of the velocity along $\xi$ axis calculated by using the WKB approximation

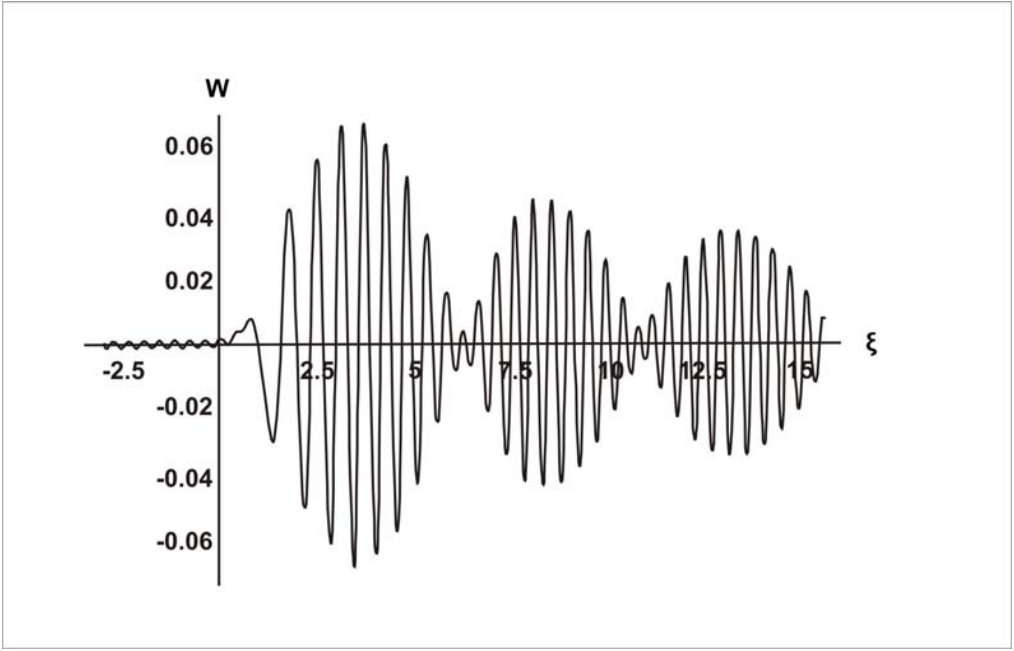

Fig.15 Vertical component $w$ of the velocity along $\xi$ axis calculated by using the uniform Airy approximation

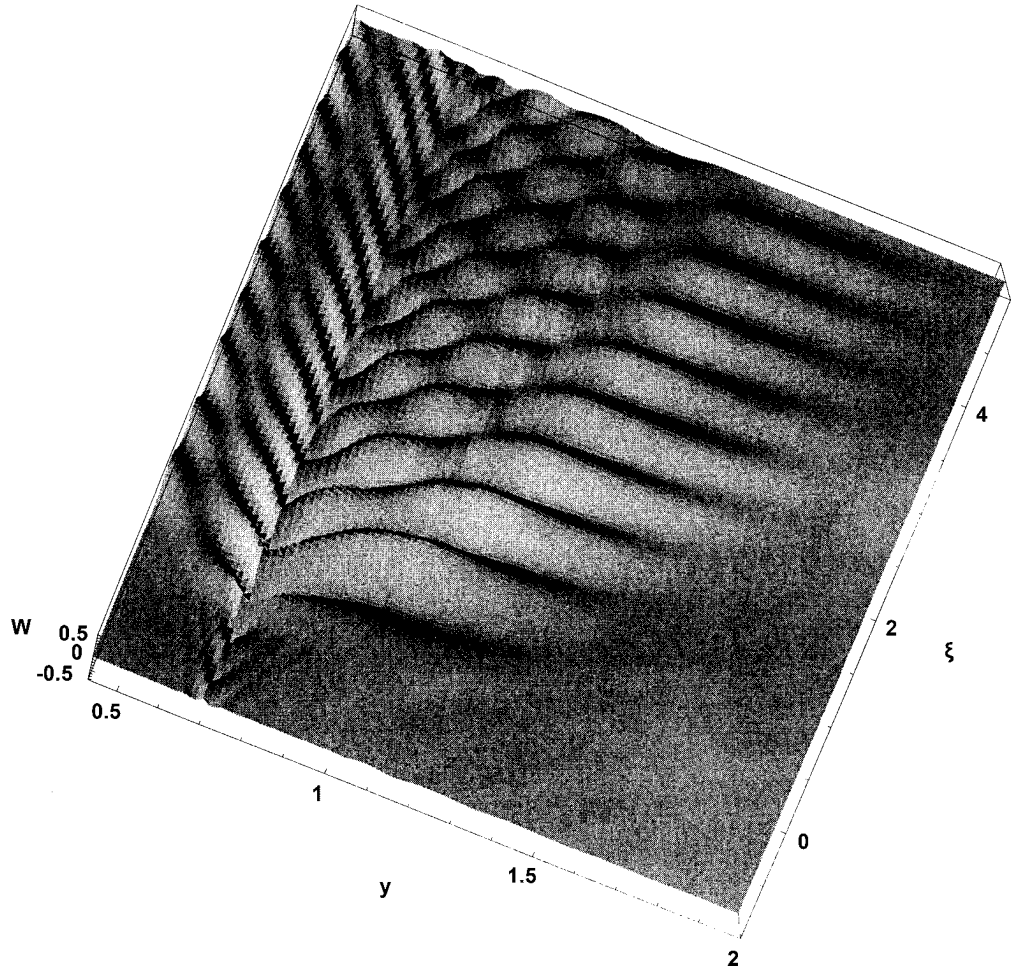

Fig.16 Vertical component $w$ of the velocity in $(\xi, y)$ plane, calculated by using of the WKB approximation.

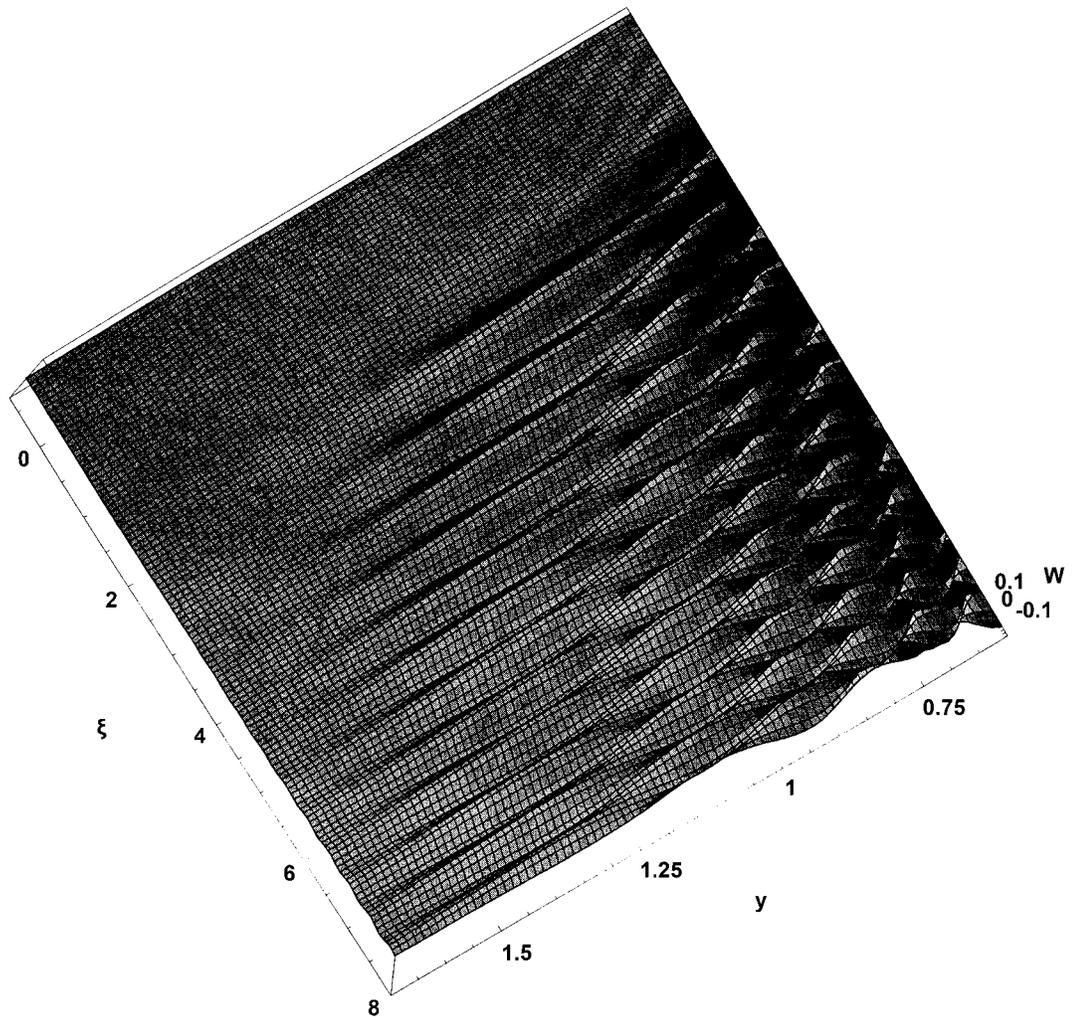

Fig.17 Vertical component $w$ of the velocity in ($\xi, y$) plane, calculated by using of the uniform Airy approximation.

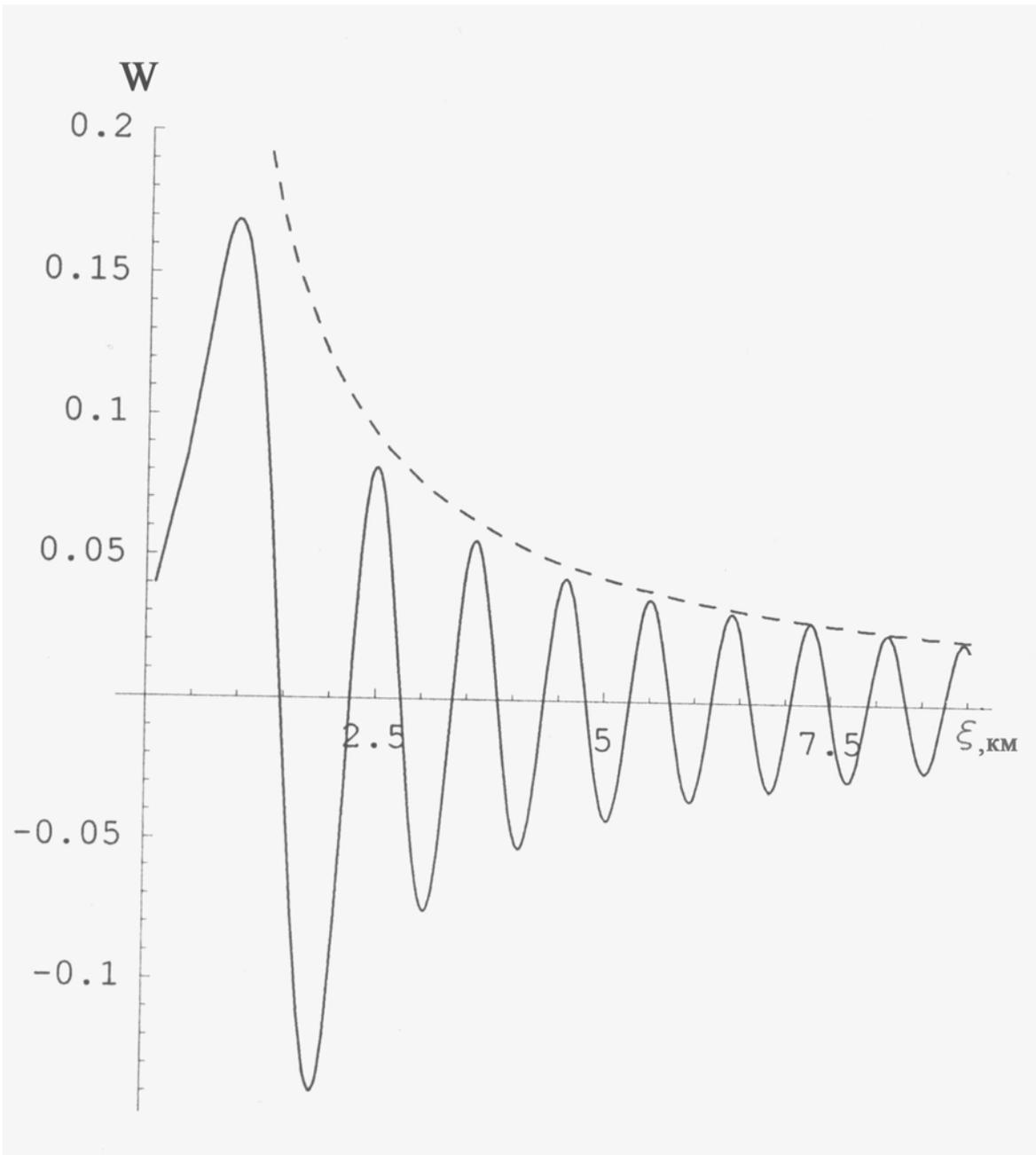

Fig.18. Vertical component $w$ of the velocity along the caustic and its approximation